\documentclass[prb,twocolumn,superscriptaddress,amsmath,amssymb,floatfix,preprintnumbers]{revtex4}
\usepackage{epsfig}
\usepackage{amsmath}
\usepackage{graphicx}
\usepackage{placeins}
\usepackage{float}
\usepackage{dcolumn}
\usepackage{color}
\usepackage{lipsum}
\usepackage{epsfig}  
\usepackage{epstopdf}
\usepackage{physics} 
\usepackage{bm}
\usepackage[all]{xy}
\usepackage{braket}
\DeclareGraphicsExtensions{.png.jpg.pdf}

\usepackage{dsfont}

\usepackage{mathtools} 
\graphicspath{{figure/}}

\usepackage{hyperref}
\hypersetup{colorlinks=true,linkcolor=blue,citecolor=blue,urlcolor=cyan}

\begin{document}
\title{Topological phase transitions in strained Lieb-Kagome lattices}
 
\author{W. P. Lima}\email{wellisson@fisica.ufc.br}
\affiliation{Departamento de F\'isica, Universidade Federal do Cear\'a, Campus do Pici, 60455-900 Fortaleza, Cear\'a, Brazil}
\author{T. F. O. Lara }\email{temersonlara@fisica.ufc.br}
\affiliation{Departamento de F\'isica, Universidade Federal do Cear\'a, Campus do Pici, 60455-900 Fortaleza, Cear\'a, Brazil}
\author{J. P. G. Nascimento} \email{joaopedro@fisica.ufc.br}
\affiliation{Departamento de F\'isica, Universidade Federal do Cear\'a, Campus do Pici, 60455-900 Fortaleza, Cear\'a, Brazil}
\author{J. Milton Pereira Jr.}\email{pereira@fisica.ufc.br}
\affiliation{Departamento de F\'isica, Universidade Federal do Cear\'a, Campus do Pici, 60455-900 Fortaleza, Cear\'a, Brazil}
\author{D. R. da Costa}\email{diego\_rabelo@fisica.ufc.br}
\affiliation{Departamento de F\'isica, Universidade Federal do Cear\'a, Campus do Pici, 60455-900 Fortaleza, Cear\'a, Brazil}
\affiliation{Department of Physics, University of Antwerp, Groenenborgerlaan 171, B-2020 Antwerp, Belgium}
	
\begin{abstract}
    Lieb and Kagome lattices exhibit two-dimensional topological insulator behavior with $\mathbb{Z}_2$ topological classification when considering spin-orbit coupling. In this study, we used a general tight-binding Hamiltonian with a morphological control parameter $\theta$ to describe the Lieb ($\theta=\pi/2$), Kagome ($\theta=2\pi/3$), and transition lattices ($\pi/2<\theta<2\pi/3$) while considering intrinsic spin-orbit (ISO) coupling. We systematically investigated the effects of shear and uniaxial strains, applied along different crystallographic directions, on the electronic spectrum of these structures. Our findings reveal that these deformations can induce topological phase transitions by modifying the structural lattice angle associated with the interconversibility process between Lieb and Kagome, the amplitude of the strain, and the magnitude of the ISO coupling. These transitions are confirmed by the evolution of Berry curvature and by changes in the Chern number when the gap closes. Additionally, by analyzing hypothetical strain scenarios in which the hopping and ISO coupling parameters remain intentionally unchanged, our results demonstrated that the strain-induced phase transitions arise from changes in the hopping and ISO coupling parameters.
\end{abstract}	
	
\maketitle
	
\section{Introduction}
    
In recent years, the investigation of topological phases of matter has emerged as an important subarea of condensed matter physics.\cite{bansilhsin2016,bernevig2013} A striking example of these phases of matter is the materials known as topological insulators, which are found to display insulating behavior in bulk, whereas their surface supports conducting states.\cite{kanecharles2008,hasankane2010,moorejoel2010} In two-dimensional (2D) topological insulators, the boundaries correspond to edges, where non-trivial topological phases give rise to gapless edge states.\cite{kanemele2005,kane2005z} In general, non-magnetic insulators with preserved time-reversal symmetry (TRS) are characterized by the $\mathbb{Z}_2$ invariant $\nu$, where an odd $\nu$ value represents a 2D topological insulator, \textit{i.e.}, quantum spin Hall insulator (QSHI), and an even $\nu$ value represents a trivial insulator.\cite{kane2005z,kanemele2005,fu2007topological,qi2008topological} Alternatively, the spin Chern number is an efficient and convenient way to distinguish different non-trivial topological states; incidentally, the $\mathbb{Z}_2$ invariant can also be calculated using the spin Chern numbers.\cite{fu2007topological,supplemental}

Topological phase transitions (TPTs) are characterized by a change in the spin Chern number calculated for topological invariants of energy bands, typically observed as the closing and reopening of the band gap.\cite{murakami2007phase, murakami2007tuning, hasankane2010} Theoretical and experimental studies have demonstrated that TPTs can be induced by manipulating the band structure through strain,\cite{agapito2013novel, liu2014manipulating, pal2014strain, qian2015topological, li2015two, kirtschig2016surface, wang2017strain, kibis2019structure, mutch2019evidence, teshome2019topological, jiang2020,nicholson2021uniaxial,bhattarai2024strain,xing2024strain} as well as through other means, such as chemical substitution, pressure, and electron correlation effects.\cite{pesin2010, wan2011topological, wray2011topological, xu2011topological, wu2013sudden} In this paper, we present calculations that demonstrate that TPTs can be driven by uniaxial, biaxial, and shear strains in 2D Lieb,\cite{weeks2010, gOLDMAN2011} transitions\cite{tony2019, Cui2019, osti_1527138, lim_2019, lang2023tilted, uchoa2024electronic} and Kagome\cite{guo2009} lattices. These lattices are QSHIs, exhibiting characteristic behavior of topological insulators with $\mathbb{Z}_2$ topological classification when the intrinsic spin-orbit (ISO) coupling is taken into account.\cite{weeks2010,titvinidze2021}

\begin{figure*}
    {\includegraphics[width=\linewidth]{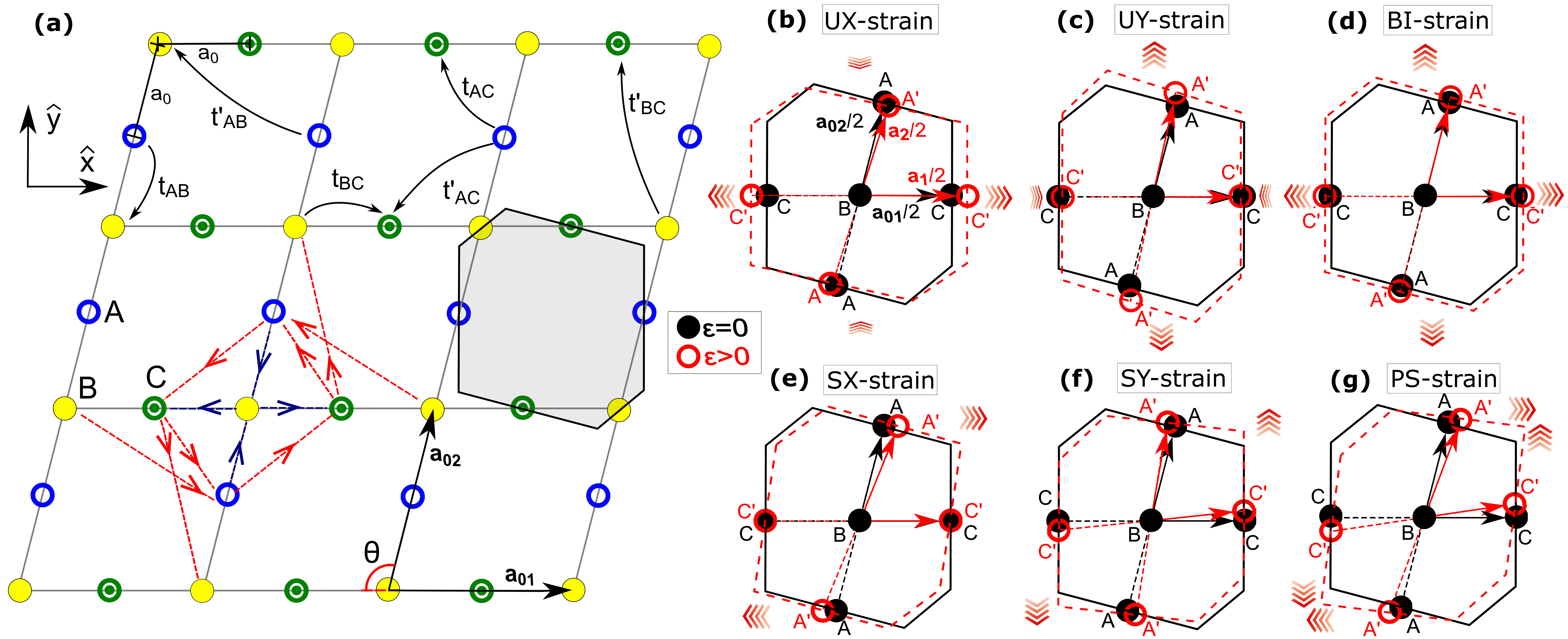}}
    \caption{(Color online) (a) Lieb-Kagome lattice characterized by the morphological parameter $\theta$. $\boldsymbol{a_{01}}$ and $\boldsymbol{a_{02}}$ are the primitive vectors of the unstrained lattice. The shaded unit cell contains three non-equivalent sites: A (blue empty circle), B (yellow filled circle), and C (green circle with a dot inside). The distance between NN sites is $a_0$, and the non-zero hopping parameters are represented by ${{t}_{AB}}$, ${{t}_{AB}'}$, ${{t}_{BC}}$, ${{t}_{BC}'}$, ${{t}_{AC}}$, and ${{t}_{AC}'}$. The ISO phase is positive ($+ i\lambda_{\langle ij\rangle}$) for spin-up electrons moving along the blue ($\lambda_{\langle ij \rangle}$) or red ($\lambda_{\langle\langle ij \rangle \rangle}$) dotted line counterclockwise, since $|\boldsymbol{e_{ij}}|=+1$. Otherwise, the ISO coupling phase is negative ($-i\lambda_{ij}$) where $|\boldsymbol{e_{ij}}|=-1$. (b)-(g) Comparison of the unit cell of the Lieb-Kagome lattice for: (i) unstrained (black solid lines - $\varepsilon=0$), and (ii) strained cases (red dashed lines - $\varepsilon>0$) subjected to: uniaxial strain along the (b) $x-$ direction (UX) and (c) $y-$ direction (UY), (d) biaxial strain (BI), (e) simple shear strain along the $x-$ direction (SX) and (f) $y-$ direction (SY), and (g) pure shear strain (PS). See Fig.~\textcolor{blue}{S1} and Secs.~\textcolor{blue}{SI} and \textcolor{blue}{SII} of the Supplemental Material.\cite{supplemental}}
    \label{FIG1}
\end{figure*}

The Lieb \cite{lieb1989} and Kagome lattices \cite{Mielke_1992} are characterized by three non-equivalent sites within their unit cells, as illustrated in Fig.~\ref{FIG1}(a). Their energy spectra display a combination of Dirac cones and flat bands,\cite{guo2009, li2016, lu_2017, Kagome_stefan, Yin_2019} as depicted in Fig.~\ref{FIG2}. These lattices are interconvertible by applying strains along the diagonal direction, allowing us to define a generic lattice known as the Lieb-Kagome lattice.\cite{tony2019, Cui2019, osti_1527138, lim_2019, lang2023tilted, uchoa2024electronic} As shown in Fig.~\ref{FIG1}(a), its 2D structure features a unit cell composed of three base sites (A, B, C), characterized by $\pi/2\leq\theta\leq2\pi/3$ (see Sec.~\textcolor{blue}{SI} of the Supplemental Material \cite{supplemental}). $\theta=\pi/2$, $\pi/2<\theta<2\pi/3$, and $\theta=2\pi/3$ correspond to the Lieb, transition, and Kagome lattices, respectively. Consequently, the Lieb-Kagome lattice can be described by a generic tight-binding (TB) Hamiltonian,\cite{tony2019, Cui2019, osti_1527138, lim_2019, lang2023tilted,lima2023} elaborated in Sec.~\textcolor{blue}{II} of the Supplemental Material.\cite{supplemental}

In Ref.~[\onlinecite{lima2023}], such Lieb-Kagome interconvertibility was employed as a methodological resource to investigate the effects of uniaxial, biaxial, and shear strains along various crystallographic directions on the energy spectrum of the Lieb and Kagome lattices. In our study here, we adopt a similar approach, but with the notable distinction that we incorporate the ISO coupling into the Lieb-Kagome lattice, aiming to elucidate the topological evolution of the Lieb, transition, and Kagome lattices under the six types of strain investigated in Ref.~[\onlinecite{lima2023}], as represented in Figs.~\ref{FIG1}(b)--\ref{FIG1}(g): uniaxial strain along $x-$ and $y-$ directions, biaxial strain, simple shear strain along $x-$ and $y-$ directions, and pure shear strain.

Previous studies have investigated TPTs in Lieb \cite{wang2016topological,chen2017spin, jiang2020} or Kagome \cite{liu2012topological,bolens2019topological, titvinidze2021, zhao2022topological,mojarro2023tilted,deng2024effect,xing2024strain} lattices, as well as in both lattices.\cite{belgeling2012, ablowitz2019, tony2019} Typically, these studies consider the inclusion of ISO coupling and have identified several methods to drive TPTs, including: (i) adjusting the strength of a real next-nearest-neighbor (NNN) hopping term,\cite{belgeling2012} (ii) incorporating Rashba spin-orbit (RSO) coupling with spin-independent and spin-dependent staggered potentials,\cite{chen2017spin} (iii) tuning the trimerization parameter by considering different hopping amplitudes on two distinct sets of lattices,\cite{bolens2019topological} (iv) decomposing the lattice into three sublattices that can move independently,\cite{ablowitz2019} (v) employing artificial gauge fields represented as spin-dependent Peierls phases,\cite{titvinidze2021} (vi) pressure,\cite{deng2024effect} and (vii) by applying mechanical strain to these lattices.\cite{jiang2020, zhao2022topological, mojarro2023tilted,xing2024strain}

Jiang \textit{et al.} \cite{jiang2020} in 2020 demonstrated TPTs in phthalocyanine-based metal-organic frameworks with a Lieb band structure via biaxial strain engineering. Similarly, Zhao \textit{et al.} \cite{zhao2022topological} in 2022 showed weaker ISO coupling-induced TPTs under uniaxial strain. Reference~[\onlinecite{mojarro2023tilted}] demonstrated theoretical TPTs in Kagome lattices under uniaxial strain, RSO coupling, and site asymmetries. More recently, Ref.~[\onlinecite{xing2024strain}] observed topological edge state modulation in 2D ferromagnetic V$_3$F$_8$ under uniaxial strain, highlighting the electronic Kagome lattice's signature transition. However, it is relevant to note that these previous studies did not employ a versatile generic TB model to explore transition lattices or conduct a comparative analysis of the effects of strain on the TPTs of Lieb and Kagome lattices. Furthermore, they primarily focused on uniaxial or biaxial strains and did not explore the full range of strain types that we have incorporated into our model. Thus, to the best of our knowledge, there is a lack of systematic studies exploring the potential TPTs driven by uniaxial and shear strains in the Lieb, transition, and Kagome lattices in the existing literature, which motivates the research discussed here. 

The paper is structured as follows: In Sec.~\ref{Sec.II}, we introduce the TB model with ISO coupling for the Lieb-Kagome lattice under strain. Sections~\ref{Sec.III} and \ref{Sec.IV} are dedicated to discussing TPTs in the unstrained and strained Lieb-Kagome lattice, respectively, presenting discussions on the closing and reopening of the band gaps, the evolution of the Berry curvature, and changes in the spin Chern numbers as evidence of TPTs. Finally, in Sec.~\ref{Sec.V}, we provide the main concluding remarks. The details of the strained Lieb-Kagome lattice (Sec.~\textcolor{blue}{SI}),\cite{chenstraingrafrevharrison,oliva2013,oliva2015,oliva_Leyva_2017,naumis2017,pereira2009UniXgraf,sena_2012,li2010,qu2014156,cocco2010} the TB model (Sec.~\textcolor{blue}{SII}),\cite{lima2022tight} comparisons with previous models for unstrained lattices (Sec.~\textcolor{blue}{SIII}), the evaluation of the Chern numbers (Sec.~\textcolor{blue}{SIV}), \cite{hor2009p,xia2009observation,hsieh2009observation,vanderbilt2018,sheng2006quantum,fukui2007topological,hatsugai1993,fukui2005chern} discussions of the negative indirect gap (Sec.~\textcolor{blue}{SVII}) and hypothetical strains (Sec.~\textcolor{blue}{SIX}), as well as additional figures (Secs.~\textcolor{blue}{SV}, \textcolor{blue}{SVI}, and \textcolor{blue}{SVIII}) can be found in the Supplemental Material.\cite{supplemental}

\section{Tight-binding model}\label{Sec.II}

The general TB Hamiltonian that describes the strained Lieb-Kagome lattices employed in this work (defined in detail in Secs.~\textcolor{blue}{SI} and \textcolor{blue}{SII} of the Supplemental Material \cite{supplemental}) reads
\begin{equation}\label{eq.1}   \hat{H}=\hat{H}_\textrm{0}+\hat{H}_\textrm{ISO},
\end{equation}
with 
\begin{subequations}\label{eq.H_HISO}
	\begin{eqnarray}
		&&\hat{H}_\textrm{0}=\sum_{i,j;\sigma}t_{ij}\hat{s}^{\dagger}_{i,\sigma}\hat{s}_{j,\sigma}+h.c.,\label{eq.H_HISO.a}\\
		&&\hat{H}_\textrm{ISO}=\hspace{-0.2cm}\sum_{ i,j;\sigma,\sigma'}\hspace{-0.2cm}i\lambda_{ij}\hat{s}^{\dagger}_{i,\sigma}\left(\textbf{{\textrm{e}}}_{ij}\cdot\boldsymbol{\sigma}_{\sigma\sigma'}\right)\hat{s}_{j,\sigma'}+h.c.\label{eq.H_HISO.b}
	\end{eqnarray}
\end{subequations}
Here, $\hat{s}_{i,\sigma}$ ($\hat{s}_{i,\sigma}^{\dagger}$) represents the annihilation (creation) operator for a particle $i$ with spin $\sigma$. The first term of Eq.~\eqref{eq.1}, $\hat{H}_\textrm{0}$ given in Eq.~\eqref{eq.H_HISO.a}, describes the hopping contributions considered in our TB model, developed in Sec.~\textcolor{blue}{SII} of the Supplemental Material.\cite{supplemental} The second term of Eq.~\eqref{eq.1}, $\hat{H}_\textrm{ISO}$ given in Eq.~\eqref{eq.H_HISO.b}, represents the ISO coupling Hamiltonian; in this term, we define the unit vector $\textbf{\textrm{e}}_{ij}=\left(\textbf{d}_{ik}\times\textbf{d}_{kj}\right)/\left|\textbf{d}_{ik}\times\textbf{d}_{kj}\right|$ in terms of the bond vectors $\textbf{d}_{ik}$ and $\textbf{d}_{kj}$, connecting the sites $i$ and $j$ through the unique intermediate site $k$. \cite{belgeling2012} Finally, $\boldsymbol{\sigma}=\left(\sigma_{x},\sigma_{y},\sigma_{z}\right)$ represents the Pauli matrices. 

The indication of the position of the sites admitted in our TB model, as well as the description of the hopping and ISO coupling parameters considered, along with the derivation of the Hamiltonian matrices in terms of these parameters, is presented in detail in Sec.~\textcolor{blue}{SII}.\cite{supplemental} Here, it is worth mentioning that in 2D lattices, hopping processes are naturally restricted to in-plane motions, making the ISO coupling effectively proportional to $\sigma_z$. The $\lambda_{ij}$'s represent the amplitudes of the ISO coupling, while $t_{ij}$'s are the hopping parameters depicted in Fig.~\ref{FIG1}(a), obeying the relation \cite{lima2022tight,lima2023}
\begin{equation}\label{hopping}
	t_{ij}=te^{-n{\left(a_{ij}/a_0-1\right)}}{a_0}/{a_{ij}}, \qquad n=8,
\end{equation}
where $t$ is the hopping parameter corresponding to the distance between nearest-neighbor (NN) sites in the unstrained Lieb-Kagome lattice, \textit{i.e.}, $a_0$, and $a_{ij}$ is the distance between the sites $i$ and $j$ of types A, B, or C in the strained lattice, calculated by $\boldsymbol{a_{ij}}=(\mathbb{I}+\overline{\epsilon})\boldsymbol{a_{0,ij}}$, with 
\begin{equation}\label{eq_epsilon}
\overline{\varepsilon}= \left( \begin{array}{ccc}
\varepsilon_{xx}-\sigma\varepsilon_{yy} & \varepsilon_{xy}\\ 
\varepsilon_{yx} & \varepsilon_{yy}-\sigma\varepsilon_{xx}\\
\end{array} \right),
\end{equation}
where $\sigma$ denotes the Poisson ratio assumed here as $\sigma=0.1$ motivated by the strained graphene case \cite{pereira2009UXgraf}, and the $\varepsilon_{ij}$ values (with $i, j = x, y$) are summarized in Table \ref{tipos_strain} for all six investigated deformations, which are depicted in Fig.~\ref{FIG1}(b)--\ref{FIG1}(g).

\begin{table}[t!]
	\caption{Strain tensor elements [Eq.~\eqref{eq_epsilon}] for each type of strain applied in the Lieb-Kagome lattices.\cite{lima2023,thiel2019shear}}\label{tipos_strain}
	\begin{ruledtabular}
		\begin{tabular}{c|c|c|c|c}
			Type of strain & $\varepsilon_{xx}$ &$\varepsilon_{xy}$ &$\varepsilon_{yx}$ & $\varepsilon_{yy}$ \\ 
			\hline             
			UX & $\varepsilon$ &0&0&0 \\
			UY & 0&0&0&$\varepsilon$ \\
			BI & $\varepsilon$ &0&0&$\varepsilon$ \\			
			SX & 0&$\varepsilon$&0&0 \\
			SY & 0&0&$\varepsilon$&0 \\
			PS & 0 &$\varepsilon$&$\varepsilon$& 0
		\end{tabular}
	\end{ruledtabular}
\end{table}

From Eq.~\eqref{hopping}, it is seen that the value of $n$ controls the magnitude of the hopping parameters. For a given distance between sites, the corresponding hopping value will decrease (increase) as the value of the $n$-parameter increases (decreases). As discussed in Ref.~[\onlinecite{lima2023}], the effects of sites more distant than NN sites are suppressed for $n\geq8$ and intensified for $n < 8$. With $n = 8$, one obtains both the nearly-flat band and a smooth transition between the Lieb and Kagome lattices,\cite{tony2019} being thus an appropriate approximation for a TB model of NN sites to describe Lieb and Kagome lattices.\cite{tony2019,lima2023} Likely, we consider for the ISO-coupling term $\lambda_{ij}$ a similar dependence, given by \cite{tony2019}  
\begin{equation}\label{eqlambdaISO}
\lambda_{i,j}=\lambda_\textrm{ISO} e^{-{\left(d_{ij}/a_0-1\right)^n}}{a_0}/{d_{ij}}, \qquad n=8,
\end{equation}
with $\lambda_\textrm{ISO}=\lambda t$, where $\lambda$ governs the strength of ISO coupling. We admit $n=8$ in agreement with the expression that governs the hopping energies [Eq.~\eqref{hopping}].

\section{Phase transition in the unstrained Lieb-Kagome lattice}\label{Sec.III}

\begin{figure}
    {\includegraphics[width=\linewidth]{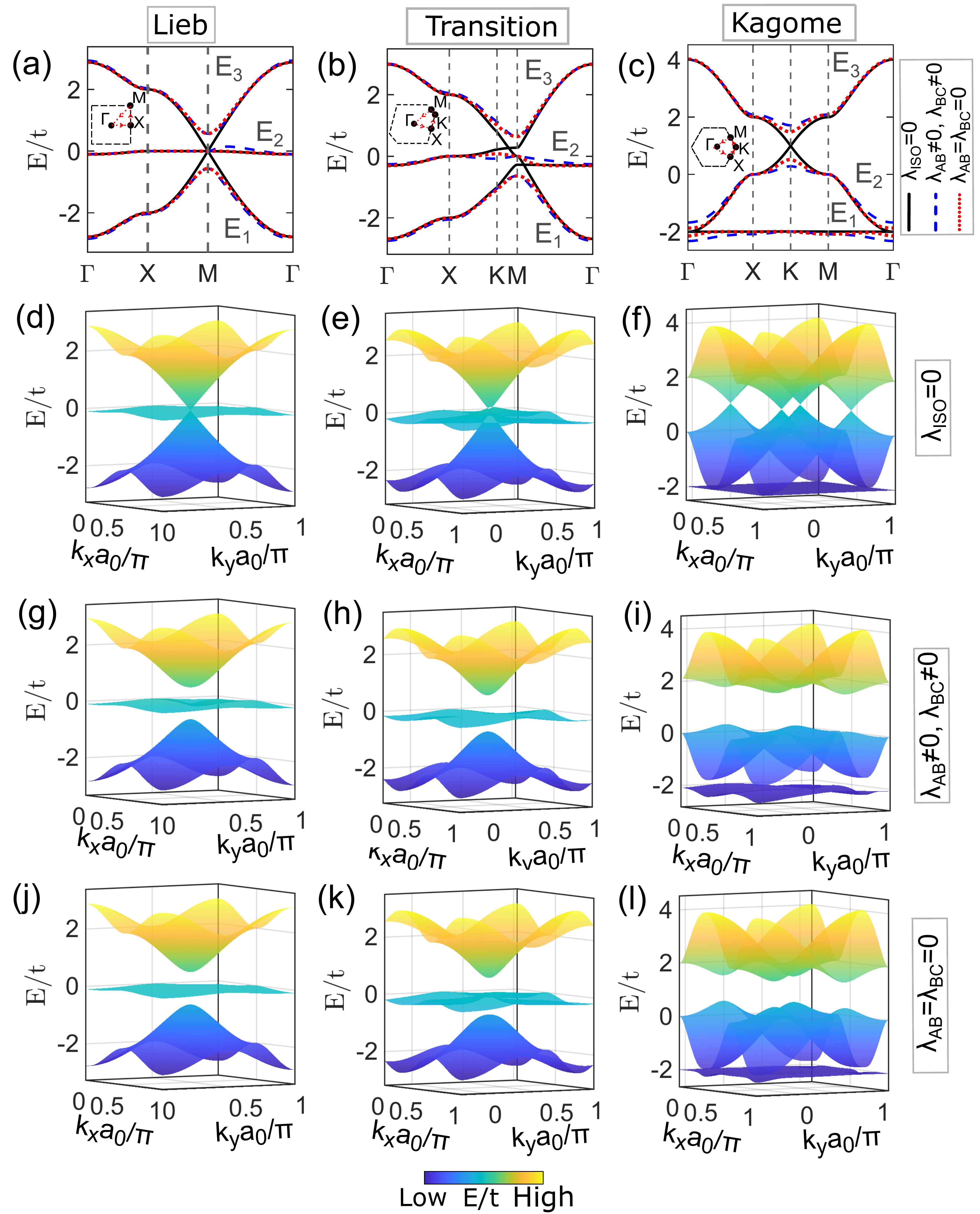}}
	\caption{(Color online) (a) Lieb, (b) transition (assuming $\theta=105^\circ$), and (c) Kagome band structures along the high-symmetry points $\boldsymbol{\Gamma}$, ${\textbf{X}}$, ${\textbf{K}}$, and ${\textbf{M}}$ of the first Brillouin Zone, as shown in the insets. Results in the absence of ISO coupling ($\lambda_{ISO}=0$) are depicted in solid black curves with the full energy spectra being shown in panels (d)--(f); in the presence of ISO coupling, for $\lambda_{AB}\neq0$ and $\lambda_{BC}\neq0$, results are shown in dashed blue curves and the 3D band structure depicted in panels (g)--(i); and results with ISO coupling but without the NN ISO coupling terms $\lambda_{AB}=\lambda_{BC}=0$ are given in dotted red curves and in panels (j)--(l) for 3D bands' format. The bottom, middle, and top bands are identified as $E_1$, $E_2$, and $E_3$, respectively.}
	\label{FIG2}
\end{figure}

\subsection{Lieb-Kagome lattice with ISO coupling}\label{subsec.IIIA}

In this section, we investigate TPTs on the unstrained Lieb-Kagome lattice. To this end, we initially analyze the effects of ISO coupling on the energy spectra of the Lieb, transition (assuming $\theta=105^\circ$), and Kagome lattices. The band structures of the Lieb, transition, and Kagome lattices in the absence of ISO coupling ($\lambda_{\text{ISO}}=0$) are shown in solid black curves in Figs.~\ref{FIG2}(a)--\ref{FIG2}(c) along the high-symmetry paths of the corresponding first Brillouin zones. For the Lieb lattice, the path follows $\boldsymbol{\Gamma} \text{--} \textbf{X} \text{--} \textbf{M} \text{--} \boldsymbol{\Gamma}$, while for both the transition and Kagome lattices, it follows $\boldsymbol{\Gamma} \text{--} \textbf{X} \text{--} \textbf{K} \text{--} \textbf{M} \text{--} \boldsymbol{\Gamma}$.\cite{supplemental} The corresponding full energy spectra are displayed in Figs.~\ref{FIG2}(d)--\ref{FIG2}\ref{FIG2}(f). For the Lieb and Kagome lattices, the well-known coexistence of Dirac and nearly-flat bands is observed.\cite{hwang2021flat} In the Lieb lattice, the nearly-flat band lies at the center of the Dirac bands, whereas in the Kagome lattice, it appears at the bottom (top) of the energy spectrum for a hopping parameter $t > 0$ ($t < 0$). This behavior arises because changing the sign of the hopping amplitude in Lieb, transitions, and Kagome lattices results in a spectrum that is exactly mirrored about $E/t = 0$: the spectrum for $t > 0$ [Figs.~\textcolor{blue}{S2}(a), \textcolor{blue}{S2}S2(c), and \textcolor{blue}{S2}S2(e)] is the vertical reflection of that for $t < 0$ [Figs.~\textcolor{blue}{S2}(b), \textcolor{blue}{S2}(d), and \textcolor{blue}{S2}(f)], as illustrated in Fig.~\textcolor{blue}{S2}.\cite{supplemental,PhysRevB.62.R6065,guo2009,tony2019} In the transition lattice, the flat band is significantly distorted, as lattice deformations break the symmetry required for its formation.\cite{Lieaau4511,tony2019}

In the absence of ISO coupling, no energy gaps appear in the spectra of the Lieb-Kagome lattices. The triply degenerate Dirac point at $\mathbf{M}$ in the Lieb lattice [Fig.~\ref{FIG2}, first column] evolves into two doubly degenerate Dirac points in the transition lattices [Fig.~\ref{FIG2}, second column], one of which shifts along the $\mathbf{M}\text{--}\boldsymbol{\Gamma}$ direction and the other along the $\mathbf{M}\text{--}\mathbf{K}$ direction, eventually forming the Kagome band structure [Fig.~\ref{FIG2}, third column].\cite{tony2019} Reference~[\onlinecite{hwang2021flat}] showed that introducing an on-site potential at the B sublattice of the Lieb lattice lifts the accidental threefold degeneracy at $\mathbf{M}$ into a twofold degeneracy that is protected by symmetry representation. Similarly, they demonstrated that the double degeneracy at the $\boldsymbol{\Gamma}$ point in the Kagome lattice is also symmetry-representation-enforced. Notably, these degeneracies can be lifted by the inclusion of ISO coupling in the Lieb-Kagome lattices while keeping the on-site energies fixed at zero,\cite{tony2019} as shown in Figs.~\ref{FIG2}(g)--\ref{FIG2}(l).

\begin{figure}
  \centering{\includegraphics[width=\linewidth]{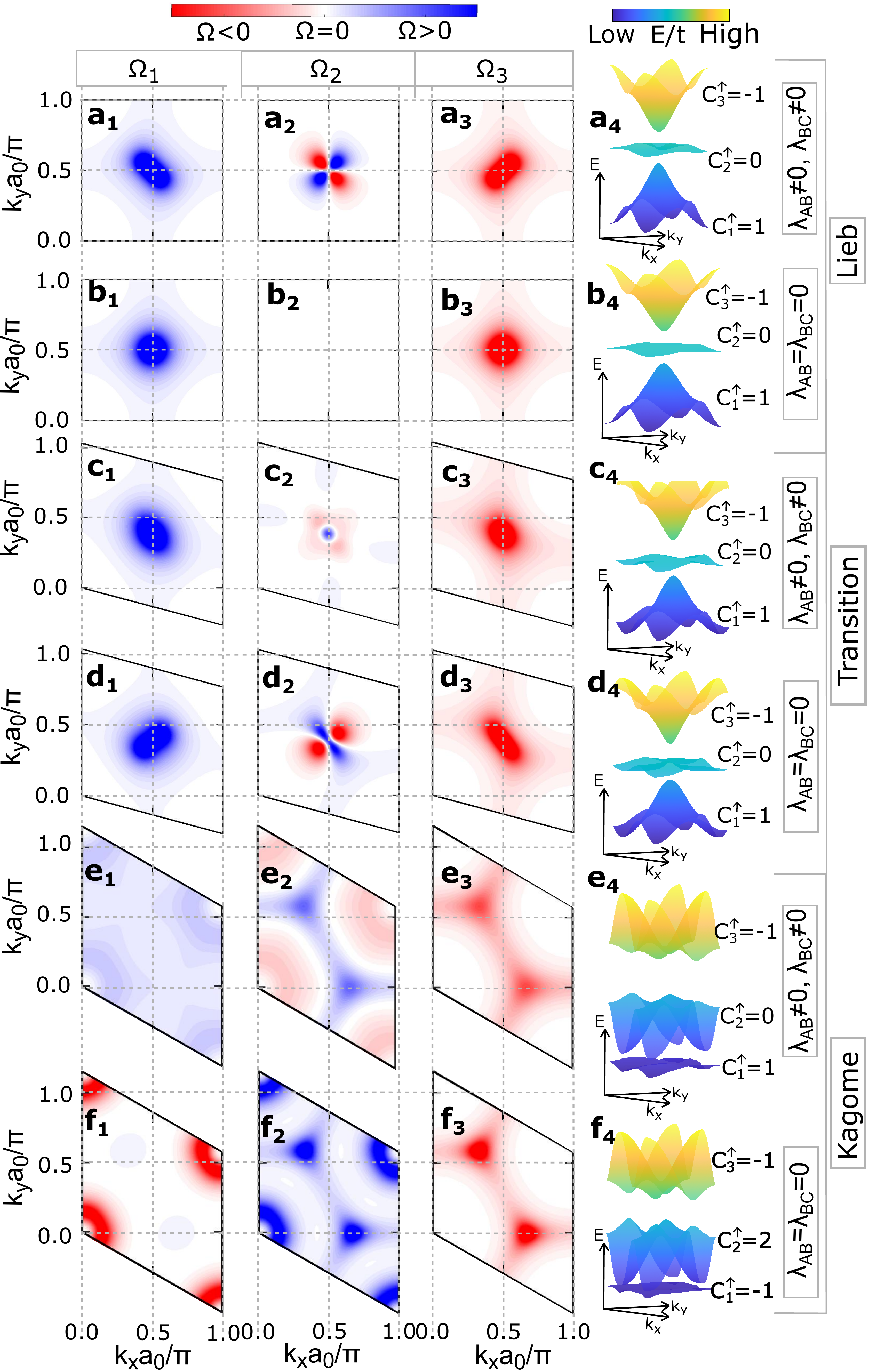}}
    \caption{(Color online) Contour plots of the Berry curvature $\Omega_1$ (panels in the first column labeled with the subscript 1), $\Omega_2$ (panels in the second column labeled with the subscript 2), and $\Omega_3$ (panels in the third column labeled with the subscript 3) corresponding to the lower, middle, and upper bands presented in the panels of the fourth column labeled with a subscript 4 for (a, b) Lieb, (c, d) transition (assuming $\theta=105^\circ$), and (e, f) Kagome lattices with (a, c, e) $\lambda_{AB}\neq0$ and  $\lambda_{BC}\neq0$, and in (b, d, f) for $\lambda_{AB}=\lambda_{BC}=0$. Parallelograms indicate the region of the reciprocal space with an area numerically equal to the first Brillouin zone of each case. Red (blue) color corresponds to $\Omega<0$ ($\Omega>0$). The spin Chern numbers $C_i^\uparrow$ are also indicated per band ($i=1,2,3$).}
    \label{FIG3}
\end{figure} 

In the presence of ISO coupling ($\lambda_{\text{ISO}} \neq 0$), two full-band gaps emerge in the Lieb-Kagome energy spectrum—$\Delta_{12}$ and $\Delta_{23}$—located between bands $E_1$ and $E_2$, and $E_2$ and $E_3$, respectively, as shown by the red and blue curves in Figs.~\ref{FIG2}(a)--\ref{FIG2}(c). In the Lieb lattice, the NN ISO coupling ($\lambda_{\langle ij \rangle}$) preserves the fourfold rotational symmetry $D_{4h}$ [see Fig.~\textcolor{blue}{S1}(b)],\cite{supplemental} and does not open a gap at the Dirac point, as evidenced in Figs.~\textcolor{blue}{S4}(a$_1$) and \textcolor{blue}{S4}(a$_2$).\cite{tony2019,hwang2021flat,supplemental} In contrast, the inclusion of NNN ISO coupling ($\lambda_{\langle\langle ij \rangle\rangle}$) breaks the $D_{4h}$ symmetry and induces the opening of a topological gap,\cite{tony2019} as illustrated in Figs.~\ref{FIG2}(g) and \ref{FIG2}(j).

In contrast to the Lieb lattice, the inclusion of the NN ISO coupling alone is sufficient to open energy gaps in the transition ($D_{2h}$) and Kagome ($D_{6h}$) lattices, as shown in Figs.~\textcolor{blue}{S4}(b$_1$), \textcolor{blue}{S4}(b$_2$), \textcolor{blue}{S4}(c$_1$), and \textcolor{blue}{S4}(c$_2$).\cite{supplemental} For completeness, all results presented in this work include the ISO coupling terms explicitly indicated in Fig.~\textcolor{blue}{S1}(c). \cite{supplemental} In fact, we systematically compare cases with and without $\lambda_{AB}$ and $\lambda_{BC}$, which are NN ISO terms for all values of $\theta$ considered; see, for instance, panels (g)--(i) and (j)--(l) of Fig.~\ref{FIG2}. Setting $\lambda_{AB} = \lambda_{BC} = 0$ removes all NN ISO contributions in the Lieb and transition lattices. This yields a hybrid Lieb model, which is not governed purely by NNN ISO coupling due to the presence of longer-range ISO contributions $\lambda_{\langle\langle\langle ij \rangle\rangle\rangle}$, shown in Fig.~\textcolor{blue}{S1}(b).\cite{supplemental} Analogously, in the Kagome lattice, setting $\lambda_{AB} = \lambda_{BC} = 0$ results in a hybrid model as well, since the $\lambda_{AC}$ term constitutes an NN ISO coupling when $\theta = 2\pi/3$ [see Fig.~\textcolor{blue}{S1}(c)].

The Berry curvatures for the lower ($\Omega_1$), middle ($\Omega_2$), and upper ($\Omega_3$) bands, as well as those associated with the lower (1/3 filling) and upper (2/3 filling) energy gaps, for the Lieb-Kagome lattices with only NN ISO coupling, are shown in Fig.~\textcolor{blue}{S4}. These results confirm that the Lieb lattice does not host a topological insulating phase when only NN ISO couplings are present, whereas the transition and Kagome lattices develop nontrivial topological phases through NN ISO coupling alone, as indicated by their nonvanishing Berry curvatures.

Throughout this work, whenever we set $\lambda_{AB} = \lambda_{BC} = 0$, we retain the resulting hybrid Lieb and Kagome models, as they provide a consistent framework aligned with the interconvertibility process. In this formulation, the ISO coupling strengths are continuously modulated according to Eq.~\eqref{eqlambdaISO}, without imposing the vanishing of any specific term strictly at the limiting angles $\theta = \pi/2$ or $\theta = 2\pi/3$. For instance, in Figs.~\ref{FIG3}(b) and \ref{FIG3}(f), the energy spectra, the gaps $\Delta_{12}$ and $\Delta_{23}$, and the associated band Chern numbers \(C^{\uparrow} = (C^{\uparrow}_{\textrm{lower}}, C^{\uparrow}_{\textrm{middle}}, C^{\uparrow}_{\textrm{upper}})\) [see Sec.~\textcolor{blue}{SIV}\cite{supplemental}] correspond to these hybrid Lieb and Kagome models and are compared to the transition model with $\lambda_{AB} = \lambda_{BC} = 0$ [Fig.~\ref{FIG3}(d)], as well as to the Lieb, transition, and Kagome lattices with $\lambda_{AB} \neq 0$ and $\lambda_{BC} \neq 0$ [Figs.~\ref{FIG3}(a), \ref{FIG3}(c), and \ref{FIG3}(e)].

The Lieb lattice exhibits a nearly flat band with zero Chern number, whereas the Kagome lattice hosts a nearly flat band with a nonzero Chern number—\textit{i.e.}, a nearly flat Chern band.\cite{hwang2021flat} Reference~[\onlinecite{hwang2021flat}] demonstrated that, in 2D flat-band systems with $C_n$ symmetry, lifting the degeneracy of a symmetry-representation-enforced band crossing via a $C_n$-symmetric perturbation can give rise to a nearly flat Chern band, as observed in the Kagome lattice. In all cases, the emergence of energy gaps due to ISO coupling signals the onset of a topologically nontrivial phase, regardless of the Chern number associated with the nearly flat band.

The evolution of the Berry curvature for the lower, middle, and upper bands of the Lieb-Kagome lattice under the ISO coupling effect is presented in Fig.~\ref{FIG3} (panels labeled with subscripts 1, 2, and 3) and in Fig.~\textcolor{blue}{S5}.\cite{supplemental} Figure~\ref{FIG3} shows the Berry curvature distributions for the Lieb, transition and Kagome lattices with and without the terms $\lambda_{AB}$ and $\lambda_{BC}$. The corresponding Berry curvatures for the 1/3 filling and 2/3 filling energy gaps in each case are presented in Fig.~\textcolor{blue}{S5}. It is observed that the nonzero Berry curvatures are mainly localized around the Dirac points, which become gapped due to the ISO coupling. The evolution of these Dirac points can thus be directly visualized through the redistribution of the Berry curvature across the three bands.\cite{tony2019,belgeling2012}

By evaluating the cases where $\lambda_{AB} = \lambda_{BC} = 0$ [Figs.~\ref{FIG3}(b), \ref{FIG3}(d), and \ref{FIG3}(f)], focusing on the evolution of the Berry curvature, one notices for Lieb lattice case that $\Omega_1$ and $\Omega_3$ exhibit circularly symmetric peaks around the \textbf{M} point with positive [Fig.~\ref{FIG3}(b$_1$)] and negative [Fig.~\ref{FIG3}(b$_3$)] magnitudes, respectively, while $\Omega_2$ is negligible due to the flatness of the flat band [Fig.~\ref{FIG3}(b$_2$)]. For transition lattice cases with $\theta = 105^\circ$, $\Omega_1$ and $\Omega_3$ remain positive and negative, respectively, similar to the Lieb case, but now display asymmetric peaks around the $\mathbf{M}$ point. $\Omega_1$ is distributed along the diagonal direction, $k_y = (\cos \theta + 1)/\sin \theta k_x$ [Fig.~\ref{FIG3}(d$_1$)], while $\Omega_3$ is along the antidiagonal direction, $k_y = (\cos \theta - 1)/\sin \theta k_x$ [Fig.~\ref{FIG3}(d$_3$)], perpendicular to $\Omega_1$. This alignment of the Berry curvature is also observed for 1/3 and 2/3 fillings, respectively, shown in Figs.~\textcolor{blue}{S5}(d$_3$) and \textcolor{blue}{S5}(d$_4$).\cite{supplemental} Interestingly, $\Omega_2$ exhibits four peaks at the $\mathbf{M}$ point, two of them being negative peaks along the diagonal direction and the other two positive peaks in the antidiagonal direction, forming Berry dipoles.\cite{zhang2018berry, battilomo2019berry} This is linked to the formation of two doubly degenerate Dirac points, located along the $\mathbf{\Gamma}$--$\mathbf{M}$ and $\mathbf{M}$--$\mathbf{K}/\mathbf{K}^\prime$ paths, originating from the triply degenerate point at the $\mathbf{M}$ point of the first Brillouin zone of the Lieb lattice when the ISO coupling is zero.\cite{tony2019} According to Fig.~\textcolor{blue}{S3},\cite{supplemental} one can confirm that the behavior of the Berry curvature contours is equivalent to that of the energy spectrum contours, particularly with respect to asymmetries around the $\mathbf{M}$ point.

The cases with $\lambda_{AB}\neq0$ and $\lambda_{BC}\neq0$ are depicted in Figs.~\ref{FIG3}(a), \ref{FIG3}(c), and \ref{FIG3}(e). Notice that the Berry curvatures of the upper and lower bands for the Lieb lattice, which were circularly symmetric around the $\mathbf{M}$ point for the case $\lambda_{AB}=\lambda_{BC}=0$,\cite{tony2019} are now asymmetric, such that the Berry curvature of the lower (upper) band resembles to the curvature of the upper (lower) band of the transition lattice with $\lambda_{AB}=\lambda_{BC}=0$. This behavior is also evidenced in the energy spectrum shown in Fig.~\textcolor{blue}{S3}(d$_1$).\cite{supplemental} Another interesting issue, observed by comparing the cases shown in Figs.~\ref{FIG3}(a) and \ref{FIG3}(b) for Lieb lattice and Figs.~\ref{FIG3}(c) and \ref{FIG3}(d) for transition lattice, is that the consideration of $\lambda_{AB}\neq0$ and $\lambda_{BC}\neq0$ originates Berry dipoles in the Berry curvature of the middle band of the Lieb spectrum [see Figs.~\ref{FIG3}(a$_2$)],\cite{battilomo2019berry} while the transition lattice no longer presenting them [see Fig.~\ref{FIG3}(c$_2$)]. Concerning that, recent studies have demonstrated the relevant role of Berry curvature dipole in driving quantum nonlinear transport phenomena, offering pathways for controlling charge currents without magnetic fields in time-reversal invariant 2D materials lacking inversion symmetry, such as transition metal dichalcogenides and twisted bilayer structures. \cite{PhysRevLett.130.016301, PhysRevB.107.205124, sinha2022berry}

\begin{figure}[ht!]
    {\includegraphics[width=\linewidth]{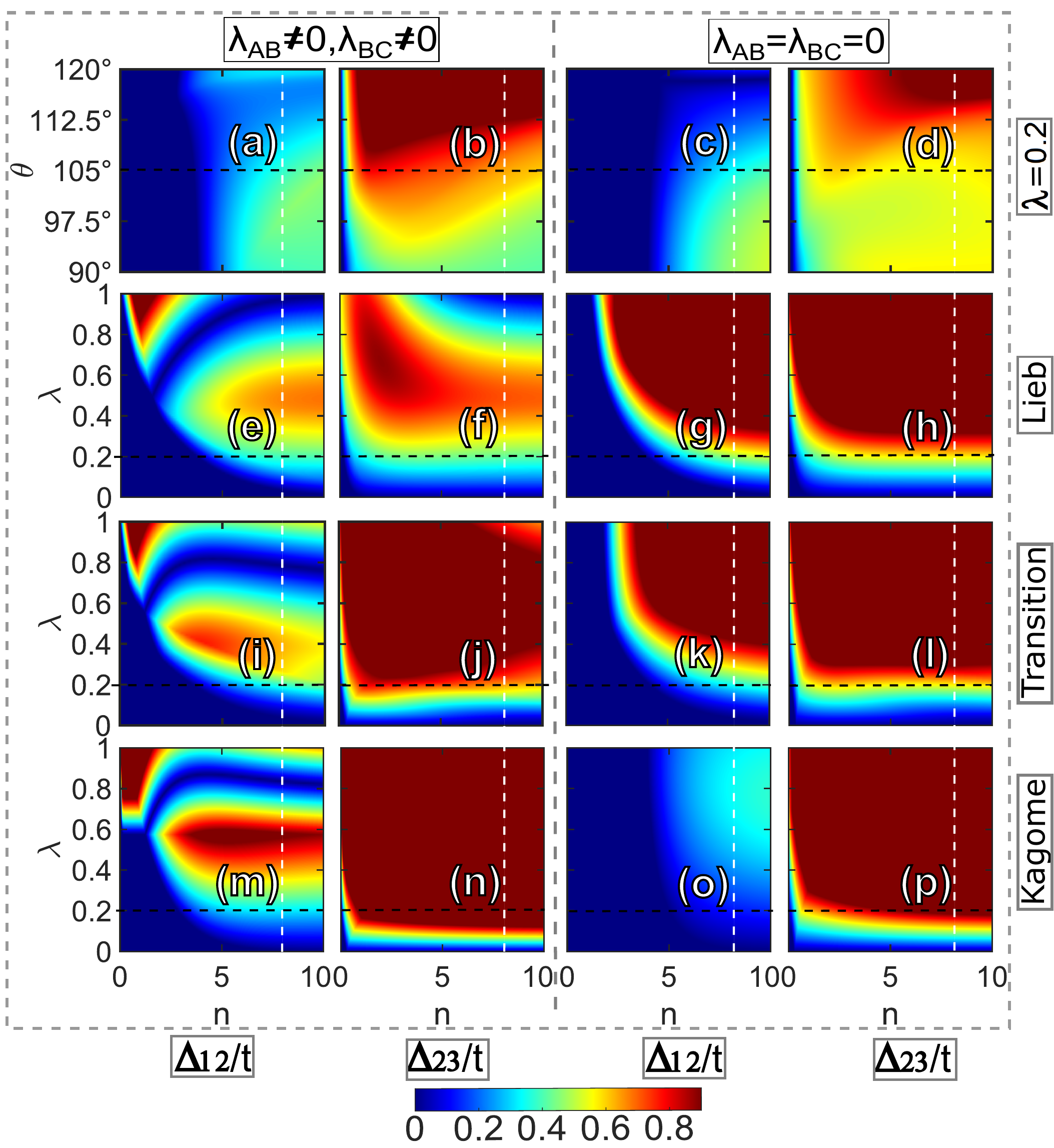}}
	\caption{(Color online) Contour plot of the full-band gap $\Delta_{12}$ ($\Delta_{23}$) between bands $E_1$ ($E_2$) and $E_2$ ($E_3$) as a function of $n$ and $\theta$ assuming $\lambda=0.2$ (1st row of panels), and as a function of $n$ and $\lambda$ for Lieb (2nd row of panels), transition (3rd row of panels), and Kagome (4th row of panels) lattices. Results are presented considering all ISO coupling parameters (1st and 2nd columns of panels), as well as assuming the following NN ISO coupling parameters equal to zero, $\lambda_{AB}=\lambda_{BC}=0$ (3rd and 4th columns of panels). The regions simultaneously exhibiting the closure of the full-band gap and the local-band gap, \textit{i.e.}, the touching of bands at a certain point \textbf{k} (shown in Fig.~\textcolor{blue}{S6})\cite{supplemental} indicate TPTs. The dashed black (white) line corresponds to the $\lambda=0.2$ ($n=8$) axis.}
\label{FIG4}
\end{figure}

By comparing the results for Lieb, transition, and Kagome shown in Figs.~\ref{FIG3}(a) \ref{FIG3}(c), and \ref{FIG3}(e), respectively, when one assumes $\lambda_{AB} \neq 0$ and $\lambda_{BC} \neq 0$, one observes that there is no change in the Chern number and in the sign of the Berry curvature between the Lieb and Kagome lattices, indicating that there are no TPTs driven by the morphological parameter $\theta$ in this configuration. The evolution of the full-band gaps as a function of the $n$ term, which controls the inclusion of NNN sites in Eq.~\eqref{hopping}, reveals that the topological phase classified by $C^{\uparrow}=(1,0,-1)$ is robust against the inclusion of NNN sites ($n<8$), provided we maintain $\lambda=0.2$ [Figs.~\ref{FIG4}(a) and \ref{FIG4}(b)]. In fact, we do not identify any closure or reopening of full-band gaps at any point in Figs.~\ref{FIG4}(a) and \ref{FIG4}(b), which means there are no TPTs. However, we note regions with full-band gaps that close and do not reopen, depicted in blue in this figure. As discussed in Sec.~\textcolor{blue}{VII} of the Supplementary Material\cite{supplemental}, in these regions the spectrum exhibits a negative indirect gap, as explicitly shown in Fig.~\textcolor{blue}{S9},\cite{belgeling2012,supplemental} such that at 1/3 filling, the system is in a semimetallic regime, characterized by partially filled bands and non-quantized spin Hall conductivity. At 2/3 filling, the system behaves as an insulator, and the spin Hall conductivity carried by the helical edge states becomes observable experimentally.\cite{belgeling2012}

For $\lambda_{AB}=\lambda_{BC}=0$, interestingly, one observes that the Berry curvature of the lower band of the Kagome lattice is negative [Fig.~\ref{FIG3}(f$_1$)], in contrast to the positive curvatures of the lower bands of the Lieb [Fig.~\ref{FIG3}(b$_1$)] and transition [Fig.~\ref{FIG3}(d$_1$)] lattices. Similarly, the middle band of the Kagome lattice becomes entirely positive [Fig.~\ref{FIG3}(f$_2$)], unlike $\theta=\pi/2$ [Fig.~\ref{FIG3}(b$_2$)] and $\theta=105^\circ$ [Fig.~\ref{FIG3}(d$_2$)] cases, which exhibit an equal volume of positive and negative phases in the first Brillouin zone, resulting in a null Chern number. However, one notes that the Kagome lattice now presents Chern number $C^{\uparrow}=(-1,2,-1)$, while the Lieb and transition ($\theta=105^\circ$) lattices show $C^{\uparrow}=(1,0,-1)$ feature. The distinct signs of the Chern number of the nearly flat band in the Kagome lattice, $C_1^\uparrow = \pm1$ [Figs.~\ref{FIG3}(e$_4$) and \ref{FIG3}(f$_4$)], depending on whether $\lambda_{AB}$ and $\lambda_{BC}$ are included or not, provide clear evidence of a TPT driven by the evolution of $\theta$, as investigated below.

\subsection{TPTs driven by $\theta$ - evolution}\label{subsec.IIIB}

\begin{figure}[t!]
    {\includegraphics[width=\linewidth]{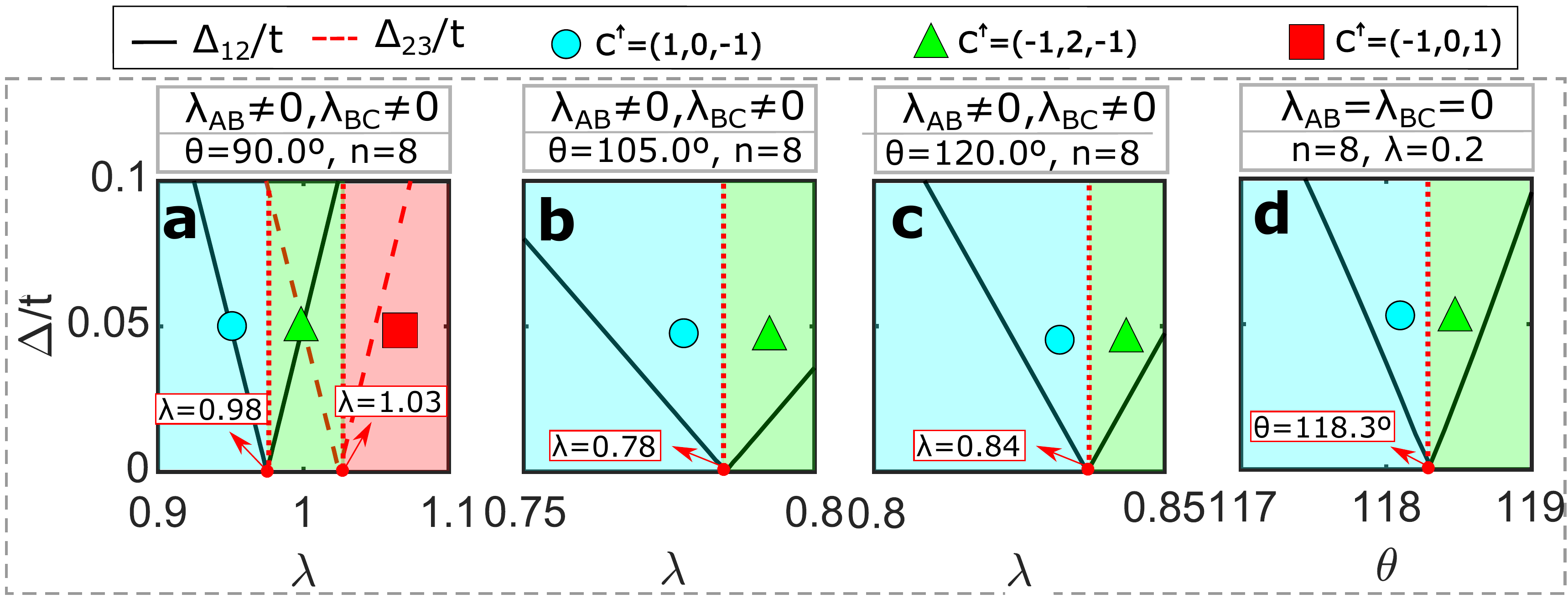}}
	\caption{(Color Online) Evolution of the full-band gaps $\Delta_{12}$ (black solid line) and $\Delta_{23}$ (red dashed line) as a function of the parameter $\lambda$ [panels (a-c)], or $\theta$ [panel (d)] assuming $\lambda_{AB}\neq0$ and $\lambda_{BC}\neq0$ [panels (a-c)] or $\lambda_{AB}=\lambda_{BC}=0$ [panels (d)], which highlights TPTs at gap closing points. Regions with distinct Chern numbers for the bands [$C^{\uparrow}=(C_{1}^{\uparrow}, C_{2}^{\uparrow}, C_{3}^{\uparrow})$] have been indicated by different colors, where $(1,0,-1)$, $(-1,2,-1)$, and $(-1,0,1)$ correspond to the blue, green, and red regions, respectively. Other cases are presented in Fig.~\textcolor{blue}{S7}.\cite{supplemental}}
\label{FIG5}
\end{figure}

The critical point of the TPT is at $\theta = 118.3^\circ$, for $\lambda_{AB}=\lambda_{BC}=0$, where band touchings are indicated by red dotted lines in the phase diagram of Fig.~\ref{FIG5}(d). Furthermore, we shaded the regime $C^{\uparrow}=(1,0,-1)$ in blue and $C^{\uparrow}=(-1,2,-1)$ in green. In all open gaps' situations, the system behaves as a QSHI, with a spin Hall conductivity of $\sigma_{\text{SH}} = \pm1$. \cite{belgeling2012} For $\theta<118.3^\circ$, the conductivities in the lower (1/3 filling) and upper (2/3 filling) gaps are the same, owing to the zero Chern number of the middle band. For $\theta>118.3^\circ$, the two bands exhibit opposite conductivities.\cite{belgeling2012} From Figs.~\ref{FIG4}(c) and \textcolor{blue}{S6}(c'), it is evident that the TPT during the interconvertibility process with $\lambda_{AB}=\lambda_{BC}=0$ remains robust for values $n\neq8$.

\begin{figure}[t!]
    \includegraphics[width=0.9\linewidth]{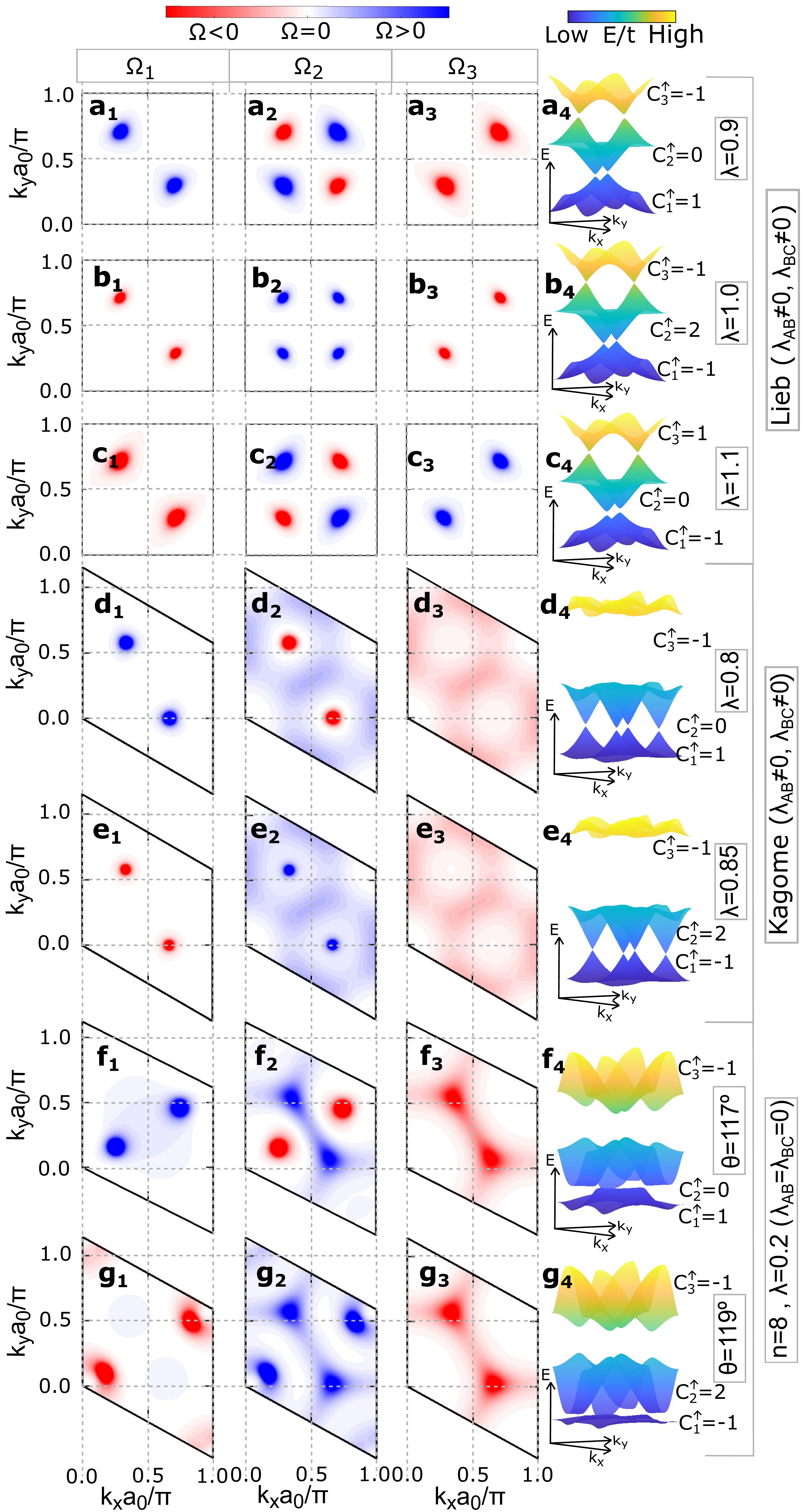}
    \caption{(Color Online) Contour plots of Berry curvature (as in Fig.~\ref{FIG3}), before and after the TPTs identified in Fig.~\ref{FIG5}. Specifically, panels (a-c), (d-e), and (f-g) correspond to the cases depicted in panels (a), (c), and (e) of Fig.~\ref{FIG5}. Other cases are presented in Fig.~\textcolor{blue}{S8}.\cite{supplemental} It is noteworthy that the TPTs cause a change in the sign of the Berry curvature.}
    \label{FIG6}
\end{figure}

Prior to this TPT [Fig.~\ref{FIG6}(f)], $\Omega_1$ shows two positive peaks along the diagonal direction. $\Omega_3$ is characterized by two negative peaks along the anti-diagonal direction. These peaks evolve from the asymmetric peak near the $\mathcal{M}$ point at $\theta = 105^\circ$, as demonstrated in Figs.~\ref{FIG3}(d$_1$) and \ref{FIG3}(d$_3$). $\Omega_2$ exhibits two positive peaks in the anti-diagonal direction and two negative peaks in the diagonal direction, as shown in Figs.~\ref{FIG3}(f$_2$), which are perpendicular to the former shown in Figs.~\ref{FIG3}(f$_1$). The TPT inverts the signs of the two peaks in $\Omega_1$ and $\Omega_3$, and in $\Omega_2$; this transformation results in the emergence of four positive peaks, eliminating the Berry dipole [Fig.~\ref{FIG6}(g)].\cite{zhang2018berry}

For the transition defined along the line $\theta < 118.3^\circ$, the lower gap closes and the system behaves as a metal at the gap-closing energy, while the upper gap remains topologically nontrivial (helical). Every TPT in the phase diagram can be understood in terms of the difference in the Chern numbers between the phases on either side of the transition.\cite{belgeling2012} Specifically, the difference in the Chern numbers is $\Delta C^{\uparrow} = (1, 0, -1) - (-1, 2, -1) = (2, -2, 0)$; that is, the change in the Chern numbers of the two touching bands is $\pm 2$. This change can be attributed to the fact that the bands touch at the $\mathbf{\Gamma}$ point and exhibit quadratic behavior in its vicinity, implying a Berry phase of $4\pi$ associated with this touching point, in accordance with Ref.~[\onlinecite{belgeling2012}].

As shown in Fig.~\ref{FIG4}(c), when compared with Fig.~\textcolor{blue}{S6}(c') in Supplementary Material\cite{supplemental}, it is also revealed that the double degeneracy at the $\mathbf{\Gamma}$ point between bands $E_1$ and $E_2$ for $\theta < 118.3^\circ$ can be lifted by either including ($n<8$) or further suppressing ($n>8$) the effects of NNN hoppings [see Eqs.~\eqref{hopping} and \eqref{eqlambdaISO}], as elucidated in the phase diagram of Fig.~\textcolor{blue}{S7}(e).\cite{supplemental} TPTs occur at $n=4.18$ with $\Delta C^{\uparrow} = (1,0,-1) - (-1,2,-1) = (2,-2,0)$ and at $n=8$ with $\Delta C^{\uparrow} = (-1,2,-1) - (1,0,-1) = (-2,2,0)$. Furthermore, the phase diagram in Fig.~\textcolor{blue}{S7}(f) of the Supplementary Material\cite{supplemental} shows that $\Delta_{12}$ can be opened while maintaining $n=8$ and evolving the intensity of the ISO coupling, such that a TPT occurs at $\lambda=0.2$ with $\Delta C^{\uparrow} = (1,0,-1) - (-1,2,-1) = (2,-2,0)$.

Interestingly, the TPT driven by $\theta$ evolution is the only phase transition observed in the case of hypothetical strains that modify the positions of the lattice sites while keeping the hopping and ISO coupling parameters constant. Figure~\textcolor{blue}{S25} of the Supplementary Material\cite{supplemental} shows that there is no closing and reopening of band gaps except at $\theta=118.3^\circ$ when $\lambda_{AB}=\lambda_{BC}=0$. Other phase transitions that are discussed in Sec~\ref{Sec.IV} are essentially due to changes in the values of the hopping parameters and ISO coupling parameters that vary with the strain parameter.

Indeed, the TPTs, as well as the non-TPTs to be discussed in Sec.~\ref{Sec.IV} are related to the evolution of energetic parameters that vary due to diagonal strains during the interconvertibility process, or due to the application of uniaxial, biaxial, simple shear, or pure strains. They are not merely a geometric effect resulting from the alteration of the positions of the Lieb-Kagome lattice sites. Thus, it is possible to find TPTs at fixed values of $\theta$ without strain, simply by altering $n$ and $\lambda$ parameters as shown in Fig.~\ref{FIG4}.

In summary, in this section we showed that the TPTs driven by interconvertibility, with a critical point at $\theta = 118.3^\circ$ for $\lambda_{AB} = \lambda_{BC} = 0$, were confirmed by the closing and reopening of energy gaps, the evolution of the Berry curvature, and the calculation of the Chern number for each energy band. All other TPTs identified in this work are likewise confirmed by the same three signatures, such as the TPTs driven by $\lambda$ and $n$ evolution discussed in the next Sec.~\ref{subsec.IIIC}.

\subsection{TPTs driven by $\lambda$ and $n$ - evolution}\label{subsec.IIIC}

In Fig.~\ref{FIG4}, we present the evolution of full-band gaps in response to the variation of $\lambda$ and $n$ parameters for Lieb, transition (assuming $\theta=105^\circ$), and Kagome lattices. By comparing these results with the one for the evolution of the local-band gaps in Fig.~\textcolor{blue}{S6}, we can highlight that the TPTs arising from the variation of $\lambda$ with $n=8$ in the phase diagrams of Figs.~\ref{FIG5}(a)-\ref{FIG5}(c), corroborated by the phase diagrams in Figs.~\textcolor{blue}{S7}(a)-\textcolor{blue}{S7}(c) which demonstrate the closing and reopening of $\Delta_{12}$ or $\Delta_{23}$ at the same values of $\lambda$ where $\Delta_{12}^\prime$ or $\Delta_{23}^\prime$ also vanish.

In the Lieb lattice, for the transition defined by the points $\lambda = 0.98$ and $\lambda = 1.03$, the lower and upper gaps respectively close, and the system behaves like a metal at the gap-closing energy; meanwhile, the other gap remains helical. The differences in the Chern numbers can be expressed as $\Delta C^{\uparrow} = (1,0,-1) - (-1,2,-1) = (2,-2,0)$ and $\Delta C^{\uparrow} = (-1,2,-1) - (-1,0,1) = (0,2,-2)$, respectively.

\begin{figure*}[t!]
    \centering
    \includegraphics[width=0.9\linewidth]{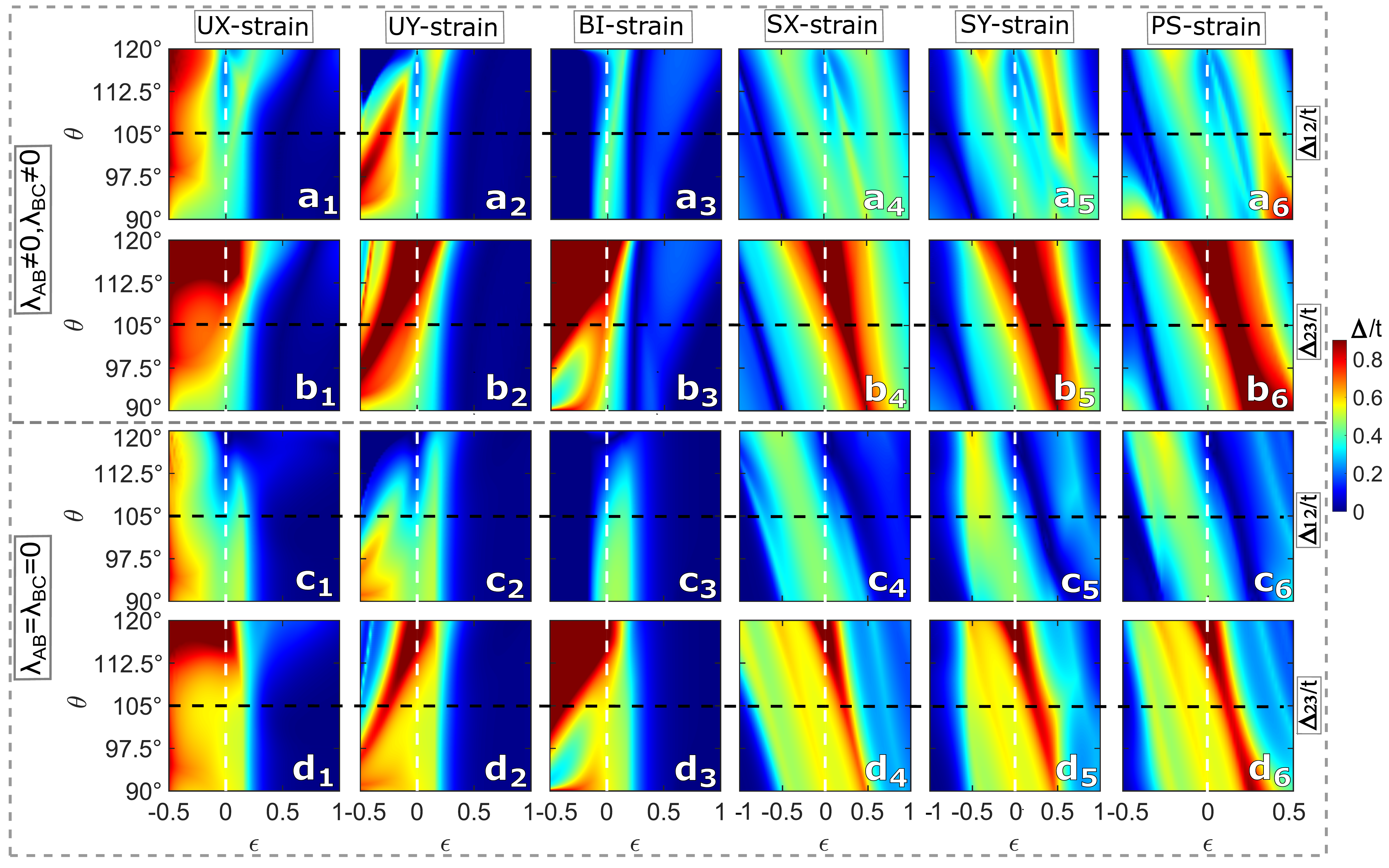}
    \caption{(Color online) Contour plot of the full-band gap similar as in Fig.~\ref{FIG4}, but now as a function of $\varepsilon$ and $\theta$, for UX-strain (1st column of panels), UY-strain (2nd column of panels), BI-strain (3rd column of panels), SX-strain (4th column of panels), SY-strain (5th column of panels), and PS-strain (6th column of panels). Results are presented considering all ISO coupling parameters (1st and 2nd rows of panels), as well as assuming $\lambda_{AB}=\lambda_{BC}=0$ (3rd and 4th rows of panels). For comparison, Fig.~\textcolor{blue}{S10} of the Supplementary Material\cite{supplemental} shows the evolution of the local-band gap in each case.}
\label{FIG7}
\end{figure*}

\begin{figure*}[t!]
    \centering
    \includegraphics[width=0.95\linewidth]{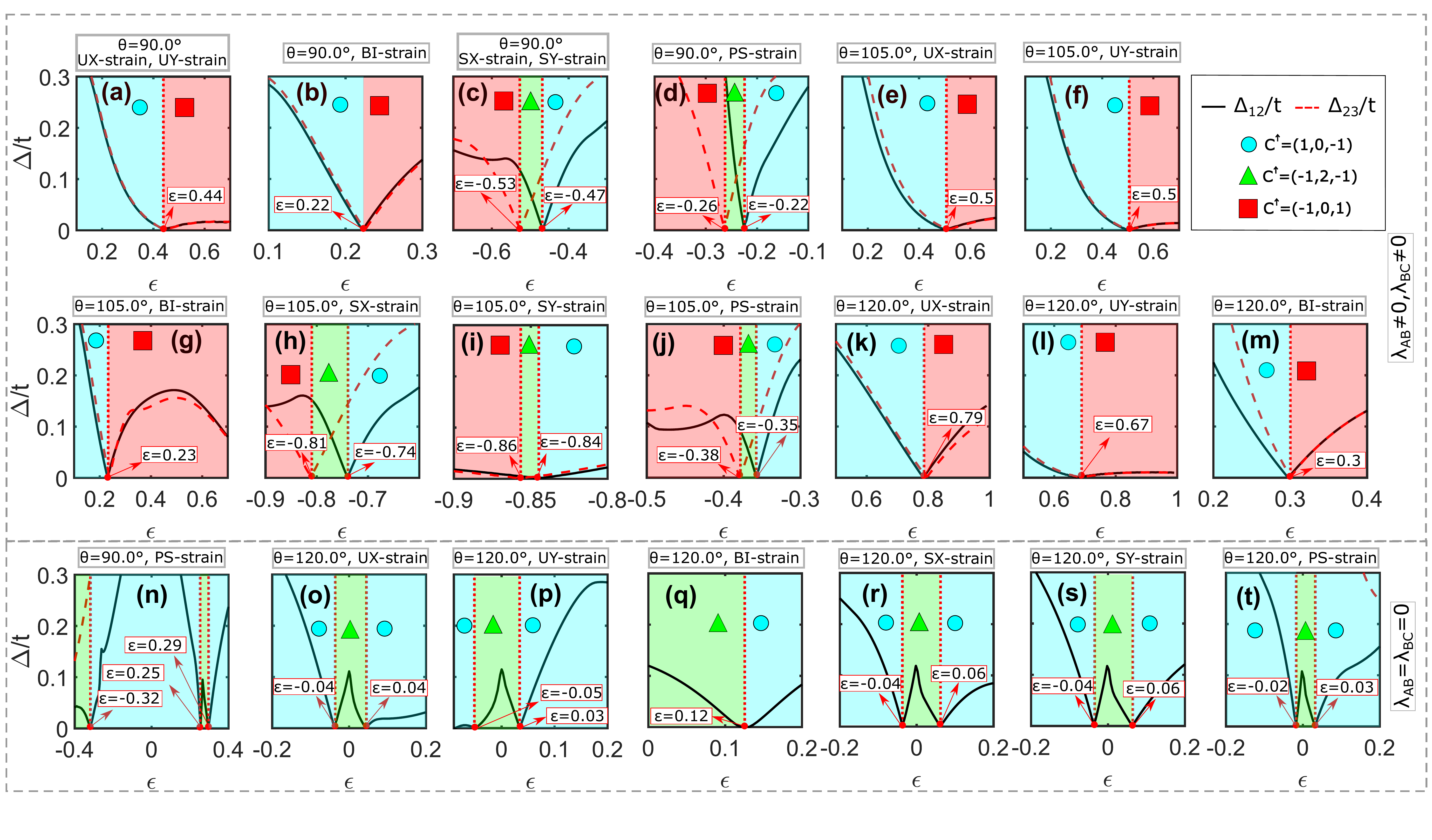}
    \caption{(Color Online) Evolution of the full-band gaps similar as in Fig.~\ref{FIG5}, but now as a function of $\varepsilon$ for the cases from Fig.~\ref{FIG7} where we identified TPTs. Figure~\textcolor{blue}{S11} of the Supplementary Material \cite{supplemental} shows the evolution of the local-band gap, confirming such TPTs.}
	\label{FIG8}
\end{figure*}

At $\lambda=0.9$, before the TPT, $\Omega_1$ ($\Omega_3$) features two positive (negative) peaks aligned in the antidiagonal (diagonal) direction, \textit{i.e.}, $k_x = -k_y$ ($k_x = k_y$) [Fig.~\ref{FIG6}(a)]. These evolved from the asymmetric peak located around the $\mathbf{M}$ point at $\lambda = 0.2$, as shown in Figs.~\ref{FIG3}(a$_1$) and \ref{FIG3}(a$_3$). $\Omega_2$ shows two positive peaks in the diagonal direction and two negative peaks in the antidiagonal direction, perpendicular to the first. The TPT at 1/3 (2/3) filling changes the signs of the two peaks of $\Omega_1$ ($\Omega_3$). Regarding $\Omega_2$, the TPT in the lower gap results in four positive peaks [Fig.~\ref{FIG6}(b)], so that it is the TPT in the upper gap that restores the dipole with the two peaks being negative (positive) in the diagonal (antidiagonal) direction [Fig.~\ref{FIG6}(c)]. \cite{zhang2018berry}

The TPTs that occur at 1/3 filling in the transition [Fig.~\ref{FIG5}(b)] and Kagome [Fig.~\ref{FIG5}(c)] lattices at the critical points $\lambda=0.78$ and $\lambda=0.84$, respectively, are analogous to the TPT in Fig.~\ref{FIG5}(d), specifically, the quantity $\Delta C^{\uparrow} = (1,0,-1) - (-1,2,-1) = (2,-2,0)$. The evolution of the Berry curvature for the transition lattice [Figs.~\textcolor{blue}{S8}(d)-\textcolor{blue}{S8}(e)]\cite{supplemental} is similar to that of the Lieb lattice [Figs.~\ref{FIG6}(a)-\ref{FIG6}(c)]. In the Kagome lattice [Figs.~\ref{FIG6}(d)-\ref{FIG6}(e)], $\Omega_3$ does not exhibit a sign change, remaining negative, but now it does not show localized peaks, analogous to $\Omega_1$ before the TPT [Fig.~\ref{FIG3}(e)]. In fact, after the TPT, the Chern nearly-flat band is $E_3$.\cite{hwang2021flat}

In summary, our results clearly indicate TPTs due exclusively to the variation of $n$ and/or $\lambda$, without the need for additional factors. With this knowledge, in the next section, we will investigate TPTs driven by the six types of strains presented in Figs.~\ref{FIG1}(b)-\ref{FIG1}(g), while maintaining $\lambda=0.2$ and $n=8$ fixed.

\section{Phase transition driven by strain}\label{Sec.IV}

In this section, we investigate the occurrence of phase transition induced by six distinct types of strains, as depicted in Figs.~\ref{FIG1}(b)-\ref{FIG1}(g). Figure~\ref{FIG7} illustrates the evolution of the full-band gaps as the angle \(\theta\) varies from \(\pi/2\) to \(2\pi/3\), in relation to the strain parameter \(\varepsilon\). For comparative analysis, the evolution of the local-band gaps is presented in Fig.~\textcolor{blue}{S10} of the Supplemental Material \cite{supplemental}. In both scenarios, while our theoretical model is not constrained by the elasticity limits of specific real materials, we impose a limitation for uniaxial and biaxial compressive strains at \(\varepsilon = -0.5\) [demonstrated in Fig.~\ref{FIG7} and panels (1), (2), and (3) of Fig.~\textcolor{blue}{S10}]. This constraint is set to prevent situations where the distance between NN sites in the strained lattice is much smaller than in the unstrained lattice (\(a_0^\prime \ll a_0\)), potentially resulting in \(\Delta_1\) and \(\Delta_2 \rightarrow \infty\). This condition is anticipated at the limit where lattice sites overlap. For simple shear strains, the overlap of site positions is not a concern, as negative \(\varepsilon\) values do not induce compression but represent shear in the opposite direction to \(\varepsilon > 0\). Thus, we allow \(-1 \leq \varepsilon \leq 1\) for these cases shown in Fig.~\ref{FIG7} and panels (4) and (5) of Fig.~\textcolor{blue}{S10}. For PS strains, we consider \(-0.5 \leq \varepsilon \leq 0.5\) [Fig.~\ref{FIG7} and panel (6) of Fig.~\textcolor{blue}{S10}], to facilitate direct comparison with simple shear cases, since \(\varepsilon^{PS} = \varepsilon^{SX}/2 = \varepsilon^{SY}/2\). \cite{thiel2019shear}

Figure~\textcolor{blue}{S11}\cite{supplemental} presents the phase diagrams with a direct comparison between the evolution of full-band gaps and local-band gaps in specific cases of Lieb lattices [see Figs.~\textcolor{blue}{S11}(a) and \textcolor{blue}{S11}(d)], the transition state (\(\theta = 105^\circ\)) [see Figs.~\textcolor{blue}{S11}(b) and \textcolor{blue}{S11}(e)], and Kagome lattices [see Figs.~\textcolor{blue}{S11}(c) and \textcolor{blue}{S11}(f)]. We explore the TPT points \(\varepsilon_c\) where closure and reopening of both full and local-band gaps at 1/3 and/or 2/3 filling are observed, attributable to the band crossing of \(E_1\) and \(E_2\) and/or \(E_2\) and \(E_3\) as a function of \(\varepsilon\).

The TPTs identified in Fig.~\textcolor{blue}{S11} are systematized in the phase diagrams of Fig.~\ref{FIG8}. As in Fig.~\ref{FIG5}, the TPT points $\varepsilon_c$ are indicated by red dotted lines, delineating the regimes $C^{\uparrow} = (1,0,-1)$ in blue, $C^{\uparrow} = (-1,2,-1)$ in green, and $C^{\uparrow} = (-1,0,1)$ in red. In all cases of open gaps, the system exhibits behavior characteristic of a QSHI, with spin Hall conductivity $\sigma_{\text{SH}} = \pm 1$. \cite{belgeling2012} When the middle band has a zero Chern number, the conductivities at 1/3 filling and 2/3 filling are the same, being opposite otherwise. \cite{belgeling2012}

\subsection{Uniaxial and biaxial strain-driven TPTs with \(\lambda_{AB} \neq 0\) and \(\lambda_{BC} \neq 0\)}\label{subsec.IVA}

\begin{figure}[t!]
    \centering
    \includegraphics[width=0.89275\linewidth]{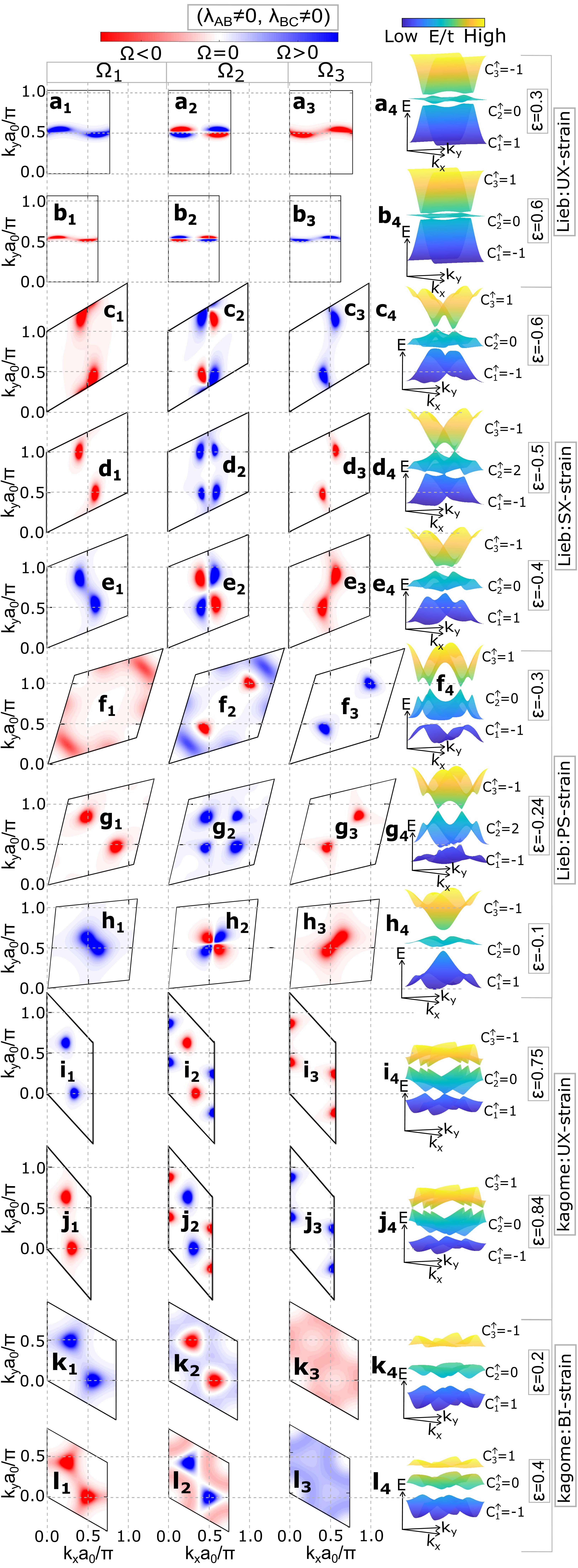}
    \caption{(Color Online) Contour plots of Berry curvature similar as in Fig.~\ref{FIG3}, for the cases before and after the TPTs identified in Fig.~\ref{FIG8} for some situations with $\lambda_{AB}\neq0$ and $\lambda_{BC}\neq0$ [see Figs.~\textcolor{blue}{S12}-\textcolor{blue}{S18} of the Supplementary Material\cite{supplemental}].}
\label{FIG9}
\end{figure}

Initially, we evaluate the TPTs driven by uniaxial and biaxial strains for \(\lambda_{AB} \neq 0\) and \(\lambda_{BC} \neq 0\). Figure~\ref{FIG8} shows that the results for Lieb [Figs.~\ref{FIG8}(a)-\ref{FIG8}(b)], transition (\(\theta = 105^\circ\)) [Figs.~\ref{FIG8}(e)-\ref{FIG8}(g)], and Kagome [Figs.~\ref{FIG8}(k)-\ref{FIG8}(m)] lattices undergo a TPT characterized by \(\Delta C^{\uparrow} = (1,0,-1) - (-1,0,1) = (2,0,-2)\). In the Lieb lattice, this TPT occurs at \(\varepsilon_c = 0.44\) when subjected to UX and UY strains [Fig.~\ref{FIG8}(a)], and at \(\varepsilon_c = 0.22\) when under BI strain [Fig.~\ref{FIG8}(b)]. In transition lattices with \(\theta = 105^\circ\), the TPT occurs at \(\varepsilon_c = 0.5\) when subjected to UX [Fig.~\ref{FIG8}(e)] and UY strains [Fig.~\ref{FIG8}(f)], and at \(\varepsilon_c = 0.23\) under BI strain [Fig.~\ref{FIG8}(g)]. In the Kagome lattice, this TPT occurs at \(\varepsilon_c = 0.79\) for the UX strain [Fig.~\ref{FIG8}(k)], \(\varepsilon_c = 0.67\) for the UY strain [Fig.~\ref{FIG8}(l)], and \(\varepsilon_c = 0.3\) for the BI strain [Fig.~\ref{FIG8}(m)]. In all cases, at \(\varepsilon = \varepsilon_c\), \(\Delta_1\) and \(\Delta_2\) close,  thereby preventing the emergence of helical states at 1/3 or 2/3 filling. These TPTs are not observed in unstrained Lieb-Kagome lattices [Fig.~\ref{FIG5}]. These TPTs are confirmed by the evolution of the Berry curvatures, as presented in Figs.~\textcolor{blue}{S14}, \textcolor{blue}{S16}, and \textcolor{blue}{S20} for the Lieb, transition, and Kagome lattices, respectively. Among these, we highlight in Figs.~\ref{FIG9}(a)-\ref{FIG9}(b), \ref{FIG9}(i)-\ref{FIG9}(j), and \ref{FIG9}(k)-\ref{FIG9}(l), the configurations in which we consider to be non-equivalent. Indeed, the other remaining cases exhibit similarities to one of these scenarios. 

The Lieb lattice under UX strain [see Figs.~\ref{FIG9}(a)-\ref{FIG9}(b)] demonstrates that \(\Omega_1^{UX}\) displays two positive peaks, while \(\Omega_3^{UX}\) exhibits two negative peaks, which reverse signs when comparing the states before and after \(\varepsilon_c\). \cite{lang2023tilted, supplemental} Meanwhile, \(\Omega_2^{UX}\) retains its dipole characteristic, swapping the signs of the two positive and two negative peaks before and after the TPT. \cite{zhang2018berry} Indeed, the Berry curvatures of Lieb lattices under UY strain [see Figs.~\textcolor{blue}{S14}(c)-\textcolor{blue}{S14}(d)] or BI strain [see Figs.~\textcolor{blue}{S14}(e)-\textcolor{blue}{S14}(f)], and of the transition lattice (\(\theta=105^\circ\)) under UX, UY, and BI strains [see Fig.~\textcolor{blue}{S16}] are analogous to those of the UX strained Lieb lattice. For Kagome lattices, the three Berry curvatures for UY strain, and \(\Omega_{1}\) for both UX and UY strains, also follow a similar pattern [see Fig.~\textcolor{blue}{S20}].

It is noteworthy that this similarity pertains to the sign configuration before and after the TPT as well as to the peak distribution of the curvature, with differences in shape and orientation. Specifically, BI-strained Lieb-Kagome lattices exhibit \(\Omega_1^{BI}\) with non-tilted peaks localized along the antidiagonal direction, defined as \(k_y = ((\textbf{b}_1 - \textbf{b}_2)_y/(\textbf{b}_1 - \textbf{b}_2)_x) k_x\), and \(\Omega_3^{BI}\) with non-tilted peaks along the diagonal direction, defined as \(k_y = ((\textbf{b}_1 + \textbf{b}_2)_y/(\textbf{b}_1 + \textbf{b}_2)_x) k_x\). Meanwhile, \(\Omega_2^{BI}\) shows two peaks each along the antidiagonal and diagonal directions. It is important to remember that \(\textbf{b}_1\) and \(\textbf{b}_2\) are dependent on \(\theta\) as demonstrated in Sec.~\textcolor{blue}{SI} of the Supplementary Material\cite{supplemental}. In contrast, UX and UY strained Lieb-Kagome lattices display tilted peaks, aligning with the direction of the reciprocal lattice vectors \(\textbf{b}_1\) and \(\textbf{b}_2\), respectively. This behavior is analogous to the type-III tilted cones observed in the energy spectrum without ISO coupling. \cite{lima2023}

The cases of \(\Omega_2\) and \(\Omega_3\) in UX and UY strained Kagome lattice exhibit the same number of peaks with the sign configurations seen in the previous cases, but with different peak distributions. In UX strained Kagome, we find four 1/2 peaks at the edges of the considered reciprocal space area [see Figs.~\ref{FIG9}(i$_{2,3}$) and ref{FIG9}(j$_{2,3}$)], as opposed to two full peaks outside the edges. In the case of BI strained Kagome, we do not identify peaks in the diagonal direction as in Lieb and transition cases, but we observe a Berry curvature distribution characteristic of a Chern nearly-flat band [see Figs.~\ref{FIG9}(k$_{2,3}$) and \ref{FIG9}(l$_{2,3}$)].

Regardless of the characteristic format of the Berry curvatures for each band, we note that the TPTs driven by uniaxial and biaxial strains for \(\lambda_{AB} \neq 0\) and \(\lambda_{BC} \neq 0\) within the admitted \(\varepsilon\) region in Fig.~\ref{FIG7} ensure the existence of a Berry curvature with both positive and negative parts calculated for the middle band, and only positive or negative curvature for the lower and upper bands. This indicates that a TPT always inverts these signs when compared with configurations before and after \(\varepsilon_c\) in Lieb-Kagome lattices.

Outside the admitted \(\varepsilon\) region in Fig.~\ref{FIG7}, we identify two TPTs in the UY strained transition lattice, which do not follow this pattern [see Fig.~\textcolor{blue}{S17}]. As observed in Fig.~\textcolor{blue}{S13}(b), one occurs at \(\varepsilon_c = -0.64\) at 1/3 filling (preserving the topological insulating phase at 2/3 filling), characterized by a Chern number change of \(\Delta C^{\uparrow} = (1,0,-1) - (-1,2,-1) = (2,-2,0)\). This change is accompanied by the alteration of the sign of the two peaks in \(\Omega_1^{UY}\), the disruption of the Berry dipole in \(\Omega_2^{UY}\) resulting in four positive peaks, and the preservation of the sign of the two peaks in \(\Omega_3^{UY}\) [see Figs.~\textcolor{blue}{S17}(b)-\textcolor{blue}{S17}(c)]. Another TPT occurs at \(\varepsilon_c = -0.68\) at 2/3 filling (maintaining the topological insulating phase at 1/3 filling), with a Chern number change of \(\Delta C^{\uparrow} = (-1,2,-1) - (-1,0,1) = (0,2,-2)\), marked by a sign change in the two peaks of \(\Omega_3^{UY}\), the restoration of the Berry dipole in \(\Omega_2^{UY}\), and the constant sign of the two peaks of \(\Omega_1^{UY}\). As we will see next, this behavior is characteristic of TPTs subjected to pure and simple shear strains within the \(\varepsilon\) range established in Fig.~\ref{FIG7}.

\subsection{Simple and pure shear strain-driven TPTs with \(\lambda_{AB} \neq 0\) and \(\lambda_{BC} \neq 0\)}\label{subsec.IVB}

Now, let us examine TPTs driven by simple and pure shear strains, when \(\lambda_{AB} \neq 0\) and \(\lambda_{BC} \neq 0\). Reducing \(\varepsilon\) to zero results in TPTs at two distinct values of \(\varepsilon_c\) in the Lieb lattice [see Figs.~\ref{FIG8}(c)-\ref{FIG8}(d)] and transition lattices [see Figs.~\ref{FIG8}(h)-\ref{FIG8}(j)] when \(\varepsilon\) falls within the range covered in Figs.~\textcolor{blue}{S11}(a$_{4-6}$) and \textcolor{blue}{S11}(b$_{4-6}$), whereas the Kagome lattice remains robust against TPTs in this scenario [see Figs.~\textcolor{blue}{S11}(c$_{4-6}$)].\cite{supplemental} The first TPT occurs at 1/3 filling, with \(\Delta C^{\uparrow} = (1,0,-1) - (-1,2,-1) = (2,-2,0)\), such that the two positive peaks in \(\Omega_1\) become negative, the characteristic Berry dipole present in \(\Omega_2^{SX}\) with two positive and two negative peaks is destroyed by the emergence of four positive peaks, and the two negative peaks of \(\Omega_3^{SX}\) maintain their sign. The second TPT occurs at 2/3 filling with \(\Delta C^{\uparrow} = (-1,2,-1) - (-1,0,1) = (0,2,-2)\), such that the two negative peaks of \(\Omega_3\) turn positive, the Berry dipole characteristic is restored in \(\Omega_2\), presenting two positive and two negative peaks (with opposite signs to the initial Berry dipole), and the sign of the two negative peaks of \(\Omega_1\) is maintained [see Figs.~\textcolor{blue}{S15} and \textcolor{blue}{S18}]. Specifically, in the Lieb lattice, TPTs occur at 1/3 filling for \(\varepsilon_c^{SX}=\varepsilon_c^{SY}=-0.47\) and at \(\varepsilon_c^{PS}=-0.22\) under PS strain. At 2/3 filling, TPTs are observed at \(\varepsilon_c^{SX}=\varepsilon_c^{SY}=-0.53\) and at \(\varepsilon_c^{PS}=-0.26\) [Fig.~\textcolor{blue}{S15}]. In the transition lattice with \(\theta=105^\circ\), TPTs at 1/3 filling occur at \(\varepsilon_c^{SX}=-0.74\), \(\varepsilon_c^{SY}=-0.84\), and \(\varepsilon_c^{PS}=-0.35\). Meanwhile, at 2/3 filling, TPTs take place at \(\varepsilon_c^{SX}=-0.81\), \(\varepsilon_c^{SY}=-0.86\), and \(\varepsilon_c^{PS}=-0.38\) [Fig.~\textcolor{blue}{S18}].

Due to the \(D_{4h}\) symmetry of the non-strained Lieb lattice, the Berry curvatures of the SY-strained Lieb lattice, \(\Omega_1^{SY}\), \(\Omega_2^{SY}\), and \(\Omega_3^{SY}\) [see Figs.~\textcolor{blue}{S15}(d)-\textcolor{blue}{S15}(f)] are analogous to the Berry curvatures of the SX-strained Lieb lattice \(\Omega_1^{SX}\), \(\Omega_2^{SX}\), and \(\Omega_3^{SX}\) [see Figs.~\textcolor{blue}{S15}(a)-\textcolor{blue}{S15}(c)]. To establish the connection between these Berry curvatures, we first apply a \(\pi/2\)-rotation using the operator \(\hat{C}_4\), followed by the application of the mirror symmetry operator \(\sigma_y\) to reflect the graph across the \(k_y\) axis. This process inversely maps points such that those with \(k_x > 0\) are mapped to \(k_x < 0\) and vice versa. Therefore, we can express the Berry curvatures for the SY case as \(\Omega_1^{SY} = \sigma_y \hat{C}_4 \Omega_1^{SX}\), \(\Omega_2^{SY} = \sigma_y \hat{C}_4 \Omega_2^{SX}\), and \(\Omega_3^{SY} = \sigma_y \hat{C}_4 \Omega_3^{SX}\). Analogously, the relation for the UY strain can be described as \(\Omega_1^{UY} = \sigma_y \hat{C}_4 \Omega_1^{UX}\), \(\Omega_2^{UY} = \sigma_y \hat{C}_4 \Omega_2^{UX}\), and \(\Omega_3^{UY} = \sigma_y \hat{C}_4 \Omega_3^{UX}\) [see Figs.~\textcolor{blue}{S14}(a)-\textcolor{blue}{S14}(d)]. Indeed, the transition lattices do not exhibit such relations between SX and SY or between UX and UY strains [see Figs.~\textcolor{blue}{S18}(a) and \textcolor{blue}{S18}(b)].

The evolution of the Berry curvatures for the PS-strained Lieb lattice, before and after the TPT at 1/3 filling [see Figs.~\ref{FIG9}(g) and \ref{FIG9}(h)], is analogous to that of the SX and SY strained Lieb lattices [see Figs.~\textcolor{blue}{S15}(a)-\textcolor{blue}{S15}(f)], with the caveat that it now exhibits mirror symmetry relative to both the diagonal and antidiagonal axes in the presented reciprocal space region. This attribute is also due to the \(D_{4h}\) symmetry of the non-strained Lieb lattice. As such, the Berry curvatures of the BI-strained Lieb lattice exhibit this symmetry [see Figs.~\textcolor{blue}{S14}(e) and \textcolor{blue}{S14}(f)], whereas those of the PS-strained transition lattice do not [see Figs.~\textcolor{blue}{S15}(g)-\textcolor{blue}{S15}(i)]. A notable difference in the SY-strained Lieb lattice occurs due to the TPT at 2/3 filling, which presents \(\Omega_{1}^{PS}\) [see Fig.~\ref{FIG9}(f$_{1}$)] with two extremely tilted peaks that differ in shape, orientation, and position from the peaks observed in the corresponding SX strain case [see Fig.~\ref{FIG9}(c$_{1}$)]. This distinction is also evident in the positive part of \(\Omega_2\) [compare Fig.~\ref{FIG9}(f$_{2}$) with Fig.~\ref{FIG9}(c$_{2}$)].

In contrast to the shear strain cases of Fig.~\ref{FIG8} with \(\lambda_{AB} \neq 0\) and \(\lambda_{BC} \neq 0\), where TPTs occurred when either \(\Delta_{12}\) or \(\Delta_{23}\) closed at different \(\varepsilon_c\), the transition lattice with PS strain exhibits a TPT when both the upper and lower gaps close and reopen at \(\varepsilon_c = 0.62\) with \(\Delta C^{\uparrow} = (1,0,-1) - (-1,0,1) = (2,0,-2)\) [see Fig.~\textcolor{blue}{S13}(c)]. Similarly, for the Kagome lattice with PS strain, a TPT occurs at \(\varepsilon_c = \pm0.49\) [see Figs.~\textcolor{blue}{S13}(d)-\textcolor{blue}{S13}(e)]. The corresponding Berry curvatures are shown in Figs.~\textcolor{blue}{S19} and \textcolor{blue}{S21}, respectively, which feature extremely tilted peaks that simply change sign as in the TPTs driven by uniaxial and biaxial strains in Figs.~\ref{FIG9}(a), \ref{FIG9}(b), \textcolor{blue}{S14}, and \textcolor{blue}{S16}.\cite{supplemental}

Notably, only the cases involving pure and simple shear in Kagome lattices exhibit mirror symmetry across $\varepsilon=0$ in the evolution of both full and local-band gaps, with respect to positive strain ($\varepsilon > 0$) and negative strain ($\varepsilon < 0$) relative to $\varepsilon = 0$, as observed in Figs.~\textcolor{blue}{S11}(c$4$-$6$). The Lieb lattice demonstrates this symmetry when considering only NNN ISO coupling, $\lambda_{\langle\langle ij \rangle\rangle}$ [see Figs.~\textcolor{blue}{S12}(d)-\textcolor{blue}{S12}(f)], represented by the red dashed lines in Fig.~\textcolor{blue}{S1}(b). This symmetry is also observed in Kagome lattices, which include only NNN ISO coupling, $\lambda_{\langle\langle ij \rangle\rangle}$ [Figs.~\textcolor{blue}{S12}(j)-\textcolor{blue}{S12}(l)] or only NN ISO coupling $\lambda_{\langle ij \rangle}$ [Figs.~\textcolor{blue}{S12}(p)-\textcolor{blue}{S12}(r)], represented by red dashed and blue dash-dotted lines, respectively, in Fig.~\textcolor{blue}{S1}(c).

Indeed, the TPTs that we found here for the Kagome lattice with \(\lambda_{AB} \neq 0\) and \(\lambda_{BC} \neq 0\) were obtained in the Kagome lattice model with only NN ISO coupling [compare the third row of panels in Fig.~\textcolor{blue}{S11} with the panels in the third row of Fig.~\textcolor{blue}{S12}]. Meanwhile, the cases of Lieb and Kagome with only NNN-ISO coupling exhibit TPTs identified in the cases with \(\lambda_{AB}=\lambda_{BC}=0\) discussed in the following subsection [compare the fourth and sixth row of panels in Fig.~\textcolor{blue}{S11} with the panels in the first and second row of Fig.~\textcolor{blue}{S12}]. The non-TPTs found in Figs.~\textcolor{blue}{S11} and \textcolor{blue}{S12} are discussed in Subsec.~\ref{subsec.Strain-driven non-topological phase transitions}.

\subsection{Strain-driven TPTs with \(\lambda_{AB}=\lambda_{BC}=0\)}\label{subsec.IVC}

\begin{figure}[t!]
    \centering
    \includegraphics[width=0.95\linewidth]{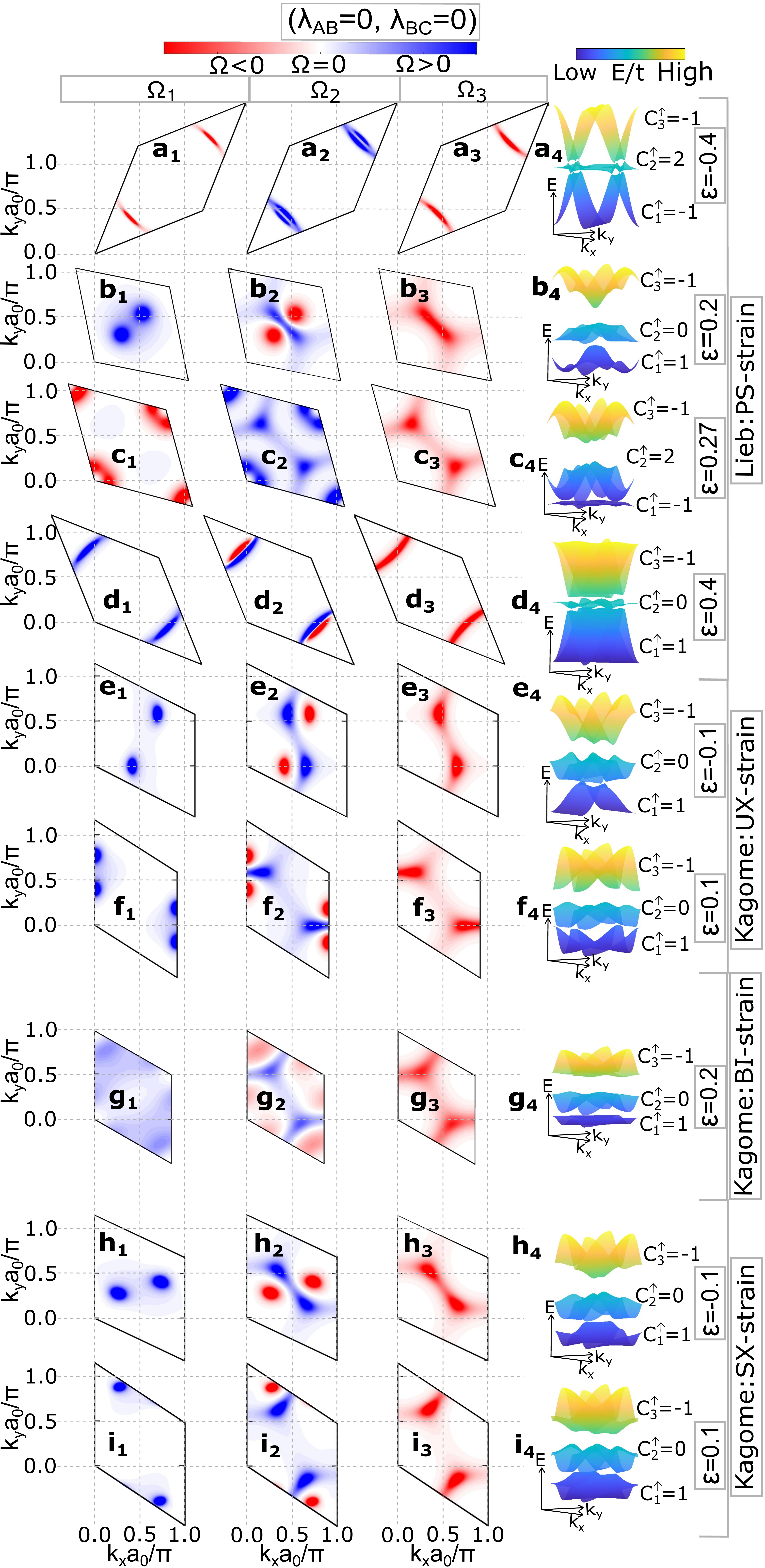}
    \caption{(Color Online) Contour plots of Berry curvature similar as in Fig.~\ref{FIG3}, for the cases before and after the TPTs identified in Fig.~\ref{FIG8} for some situations with $\lambda_{AB}=\lambda_{BC}=0$ [see Figs.~\textcolor{blue}{S19}-\textcolor{blue}{S22} fo the Supplementary Material\cite{supplemental}].}
    \label{FIG10}
\end{figure}

When \(\lambda_{AB} = \lambda_{BC} = 0\), Figs.~\ref{FIG8} (n)-\ref{FIG8}(t) demonstrate that TPTs occur exclusively at 1/3 filling within the \(\varepsilon\) range covered in Fig.~\textcolor{blue}{S11}. This is observed in the case of PS-strained Lieb [Fig.~\ref{FIG8}(n)] and Kagome lattices subjected to six types of strain [Figs.~\ref{FIG8}(o)-\ref{FIG8}(t)]. In these situations, we only identify TPTs with \(\Delta C^{\uparrow} = (1,0,-1) - (-1,2,-1) = (2,-2,0)\) or \(\Delta C^{\uparrow} = (-1,2,-1) - (1,0,-1) = (-2,2,0)\), resulting in a change of sign in the Berry curvatures $\Omega_1$ and $\Omega_2$, while $\Omega_3$ remains negative, as depicted in Fig.~\ref{FIG10} (panels with subscript 3).

Indeed, the Lieb lattice with $\lambda_{AB} = \lambda_{BC} = 0$, when subjected to uniaxial strains along the UX-strain and UY-strain directions, shows a gradual reduction in the energy gap as $\varepsilon$ increases from zero to 1. However, a complete closure of the gap does not occur within the examined parameter range [see Figs.~\textcolor{blue}{S11}(d$_1$) and \textcolor{blue}{S11}(d$_2$)]. Conversely, in the case of BI-strain, a gap closure is observed at a critical strain value of $\varepsilon = 0.68$. Beyond this critical point, the system exhibits a closed gap for $\varepsilon > 0.68$, indicating a transition to a metallic phase [see Fig.~\textcolor{blue}{S11}(d$_3$)]. Notably, neither the SX-strain [Fig.~\textcolor{blue}{S11}(d$_4$)] nor the SY-strain [Fig.~\textcolor{blue}{S11}(d$_5$)] lead to gap closure at any value of $\varepsilon$.

In the PS-strained Lieb lattice, as shown in Fig.~\ref{FIG8}(n), increasing \(\varepsilon\) from \(\varepsilon = 0\) to \(\varepsilon > 0\) leads to a first TPT at \(\varepsilon = 0.25\), with \(\Delta C^{\uparrow} = (1,0,-1) - (-1,2,-1) = (2,-2,0)\), and a second TPT at \(\varepsilon = 0.29\), with \(\Delta C^{\uparrow} = (-1,2,-1) - (1,0,-1) = (-2,2,0)\). In contrast, when \(\varepsilon\) decreases from \(\varepsilon = 0\) to \(\varepsilon < 0\), a single TPT occurs at \(\varepsilon = -0.32\), with \(\Delta C^{\uparrow} = (1,0,-1) - (-1,2,-1) = (2,-2,0)\). Interestingly, when only the NNN ISO coupling is considered in the PS-strained Lieb lattice, TPTs are observed at \(\varepsilon = \pm 0.32\) [Fig.~\textcolor{blue}{S12}(f)]. This indicates that the TPTs at \(\varepsilon = 0.25\), \(\varepsilon = 0.29\), and \(\varepsilon = 0.32\) in the model with \(\lambda_{AB} = \lambda_{BC} = 0\) arise from the effects of ISO coupling beyond NNN connections, which are nonzero in Fig.~\ref{FIG8}(n) but vanish in Fig.~\textcolor{blue}{S12}(f), while the NN ISO couplings are zero in both cases.

The evolution of the Berry curvatures presented in Fig.~\ref{FIG10}(b) shows that applying PS-strain with \(0 < \varepsilon < 0.2\) does not result in TPTs, but causes the splitting of the positive (negative) peak of $\Omega_1$ ($\Omega_3$) into two peaks distributed along the diagonal (antidiagonal) direction, when compared to the Berry curvatures of the non-strained lattice [Fig.~\ref{FIG3}(b)]. Furthermore, $\Omega_2$ exhibits two negative peaks in the diagonal direction and two positive peaks in the antidiagonal direction. The TPT at \(\varepsilon = 0.25\) leads to a configuration of Berry curvatures [Fig.~\ref{FIG10}(c)] analogous to that of the non-strained Kagome lattice [Fig.~\ref{FIG3}(f)], with $\Omega_1^{PS}$ showing negative peaks localized at the $\Gamma$ points and $\Omega_2^{PS}$ being entirely positive. Subsequently, the TPT at \(\varepsilon = 0.29\) reverses the signs of the peaks of $\Omega_1^{PS}$ and $\Omega_2^{PS}$ back to positive and mixed (two positive and two negative), respectively, as observed for $\varepsilon=0.4$ in Fig.~\ref{FIG10}(d). However, now they are significantly tilted compared to the case at $\varepsilon=0.2$ in Fig.~\ref{FIG10}(b). On the other hand, the TPT at \(\varepsilon = -0.32\) results in Berry curvatures as shown in Fig.~\ref{FIG10}(a), which exhibit highly tilted peaks similar to those in Fig.~\ref{FIG10}(d), but with a sign configuration observed in Fig.~\ref{FIG10}(c), due to similar Chern number per band.

Turning our attention to the transition lattice with a configuration angle of $\theta=105^\circ$ and ${\lambda}_{AB}={\lambda}_{BC}=0$, we find that UX-strain does not result in gap closure [Fig.~\textcolor{blue}{S11}(e$_1$)]. Similarly, UY-strain does not induce the closure of $\Delta_{12}$ within the considered $\varepsilon$ range [Fig.~\textcolor{blue}{S11}(e$_2$)]. When considering the biaxial BI-strain [Fig.~\textcolor{blue}{S11}(e$_3$)], we observe a reduction in the gap for increasingly positive values of $\varepsilon$, though the gap closure does not occur definitively. Both SX-strain [Fig.~\textcolor{blue}{S11}(e$_4$)] and SY-strain [Fig.~\textcolor{blue}{S11}(e$_5$)] do not lead to gap closure. Notably, under SY-strain at $\varepsilon=0.3$, the gap $\Delta_{12}$ reaches a minimum of $\Delta_{12}/t=0.01$, while maintaining consistent Chern numbers before and after this value. Similarly, the PS-strain [Fig.~\textcolor{blue}{S11}(e$_6$)] case does not result in gap closure within the examined range.

Indeed, in the case of transition lattices, TPTs were only observed beyond the considered range of $\varepsilon$, as shown in Fig.~\textcolor{blue}{S13}(a) of the Supplementary Material\cite{supplemental}, presenting TPTs at $\varepsilon_c=-0.6$ at 1/3 filling with \(\Delta C^{\uparrow} = (1,0,-1) - (-1,2,-1) = (2,-2,0)\) and at $\varepsilon_c=-0.64$ at 2/3 filling with \(\Delta C^{\uparrow} = (-1,2,-1) - (-1,0,1) = (0,2,-2)\), analogous to the TPTs in the corresponding case with \(\lambda_{AB} \neq 0\) and \(\lambda_{BC} \neq 0\) already discussed in the previous subsection [see Fig.~\textcolor{blue}{S13}(b)]. The evolution of the Berry curvatures corresponding to these TPTs shown in Fig.~\textcolor{blue}{S22} is also analogous to those already discussed in Fig.~\textcolor{blue}{S17}, for the equivalent case mentioned.

The Kagome lattice is highly sensitive to TPTs with \(\Delta C^{\uparrow} = (-1,2,-1) - (1,0,-1) = (-2,2,0)\), when varying \(\varepsilon\) from its initial value of zero, with \(\lambda_{AB} = \lambda_{BC} = 0\).  Specifically for UX-strain, TPTs occur at \(\varepsilon_c^{UX} = \pm0.04\) [see Fig.~\ref{FIG8}(o)]. In the case of UY-strain, transitions are observed at \(\varepsilon_c^{UY} = 0.03\) and \(\varepsilon_c^{UY} = -0.05\) [see Fig.~\ref{FIG8}(p)], while for BI-strain, a single TPT manifests at \(\varepsilon_c^{BI} = 0.12\) [see Fig.~\ref{FIG8}(q)]. SX and SY strains induce TPTs at \(\varepsilon_c^{SX} = 0.06\) and \(\varepsilon_c^{SY} = -0.04\) respectively [see Figs.~\ref{FIG8}(r) and \ref{FIG8}(s)], and PS-strain leads to transitions at \(\varepsilon_c^{PS} = 0.03\) and \(\varepsilon_c^{PS} = -0.02\) [see Fig.~\ref{FIG8}(t)].

The TPTs observed in uniaxially strained or simply and purely sheared Kagome lattices also occur when admitting only NNN ISO coupling, but at different $\varepsilon_c$ values. Specifically, at $\varepsilon_c^{UX}=-0.15$ and $\varepsilon_c^{UX}=0.11$ [Fig.~\textcolor{blue}{S12}(g)], at $\varepsilon_c^{UY}=\pm0.12$ [Fig.~\textcolor{blue}{S12}(h)], at $\varepsilon_c^{SX}=\varepsilon_c^{SY}=\pm0.14$ [Figs.~\textcolor{blue}{S12}(j) and \textcolor{blue}{S12}(k)], and at $\varepsilon_c^{PS}=\pm0.7$ [Fig.~\textcolor{blue}{S12}(l)]. Intriguingly, the TPT identified at 1/3 filling in the BI-strained Kagome lattice ceases to occur when all NN ISO couplings are nullified [Fig.~\textcolor{blue}{S12}(i)]. We conclude that this is due to the NN ISO coupling parameter $\lambda_{AC}$, which remains nonzero in Fig.~\ref{FIG8}(q).

The TPTs driven by uniaxial and biaxial strain are confirmed by the Berry curvatures shown in Fig.~\textcolor{blue}{S23}. The UX-strain stretch ($\epsilon>0$) causes negative peaks in $\Omega_1$ at the $\Gamma$ point in the non-strained Kagome lattice [see Fig.~\ref{FIG3}(f$_1$)], transforming it into four positive half-peaks. These half-peaks are distributed as two per face in the admitted reciprocal space area, along the $\mathbf{b_2}$ direction [Fig.~\ref{FIG10}(f$_1$)]. Similarly, $\Omega_2$ exhibits the same four half-peaks but, being negatives, in addition to two positive peaks outside the faces [Fig.~\ref{FIG10}(f$_2$)]. Conversely, the compressed UX-strain ($\epsilon<0$) leads to two positive (negative) peaks outside the faces in $\Omega_1$ ($\Omega_2$), diverging from the four half-peak pattern. This behavior is reversed in the case of the UY-strain case, such that the Berry curvatures for stretched UX-strain correspond to those for compressed UY-strain, and vice versa. This can be observed by comparing a$_1$ $\longleftrightarrow$ g$_1$, a$_2$ $\longleftrightarrow$ g$_2$, a$_3$ $\longleftrightarrow$ g$_3$ and c$_1$ $\longleftrightarrow$ e$_1$, c$_2$ $\longleftrightarrow$ e$_2$, c$_3$ $\longleftrightarrow$ e$_3$ in Fig.~\textcolor{blue}{S23} of the Supplementary Material \cite{supplemental}.

Interestingly, the TPT in BI-strained Kagome lattice with $\lambda_{AB} = \lambda_{BC} = 0$ results in Berry curvatures similar to those of the non-strained Kagome lattice with $\lambda_{AB} \neq 0$ and $\lambda_{BC} \neq 0$. This can be observed by comparing Figs.~\ref{FIG10}(g$_1$), \ref{FIG10}(g$_2$), and \ref{FIG10}(g$_3$) with those in Figs.~\ref{FIG3}(e$_1$), \ref{FIG3}(e$_2$), and \ref{FIG3}(e$_3$), respectively. For the SX-strained Kagome lattice, the TPT occurring at \(\varepsilon<0\) results in Berry curvatures [Fig.~\ref{FIG10}(h)] analogous to those observed in the compressed UX-strained Kagome lattice [Fig.~\ref{FIG10}(e)]. Conversely, the TPT at \(\varepsilon>0\) differs from those in the stretch UX-strained Kagome, once $\Omega_1^{SX}$ and $\Omega_2^{SX}$ display two complete peaks touching both the upper and lower faces in the considered reciprocal space region [Fig.~\ref{FIG10}(i$_{1,2}$)], instead of four half-peaks on the right and left faces, as seen in $\Omega_1^{UX}$ and $\Omega_2^{UX}$ in Fig.~\ref{FIG10}(f$_{1,2}$). The SY- and PS-strained Kagome cases are analogous to the SX-strained Kagome as shown in Fig.~\textcolor{blue}{S24} of the Supplementary Material \cite{supplemental}.

We emphasize that usually one needs strains of at least $\varepsilon=0.2$ in order to close the band gap. Furthermore, the cases with $\lambda_{AB}\neq0$ and $\lambda_{BC}\neq0$ show more difficulty in closing the gap for the Lieb lattice when compared to the case $\lambda_{AB}=\lambda_{BC}=0$. Also, smaller strains are sufficient to close (increase) the gap with extension (compression) cases than when one has compression (extension).  

\subsection{Strain-driven non-topological phase transitions}\label{subsec.Strain-driven non-topological phase transitions}\label{subsec.IVD}

We did not exclusively identify TPTs when applying strain to the Lieb-Kagome lattices; non-TPTs were also observed. When both band gaps, $\Delta_{12}$ and $\Delta_{23}$, as well as $\Delta_{12}^\prime$ and $\Delta_{23}^\prime$ close at $\varepsilon_c$ and do not reopen, the system transitions to a metallic state, thereby losing its QSHI characteristics [yellow regions in Figs.~\textcolor{blue}{S11} and ~\textcolor{blue}{S12}]. This occurs in the BI-strained Lieb lattice for $\varepsilon > 0.64$, regardless of the inclusion of $\lambda_{AB}$ and $\lambda_{BC}$ terms [Figs.~\textcolor{blue}{S11}(a$_3$) and \textcolor{blue}{S11}(d$_3$)], and remains the case when considering only NNN ISO coupling [Fig.~\textcolor{blue}{S12}(c)]. Interestingly, this phase is also observed at $\varepsilon > 0.3$ in the BI-strained Kagome lattice when only NNN ISO coupling is admitted in the model [Fig.~\textcolor{blue}{S12}(i)]. When the NN ISO coupling term $\lambda_{AC}$ is introduced in this case, instead of a metallic phase, we have a topological insulator phase due to the TPT at $\varepsilon_c^{BI}=0.12$ discussed in the previous subsection [see Fig.~\ref{FIG8}(q)].

Conversely, when only the full-band gap $\Delta_{12}$ closes while the corresponding local-band gap $\Delta_{12}^\prime$ remains nonzero, the spectrum starts exhibiting a negative indirect gap at 1/3 filling [orange regions in Figs.~\textcolor{blue}{S11} and ~\textcolor{blue}{S12}]. This phenomenon classifies the bulk as a semimetal, wherein the QSHI state is only present in the upper gap. In this case, either the lower and middle bands are partially filled, precluding the formation of helical edge states at 1/3 filling. For a detailed discussion, see Sec.~\textcolor{blue}{VI} in the Supplemental Material \cite{supplemental}.

When $\lambda_{AB} \neq 0$ and $\lambda_{BC} \neq 0$, a semimetallic phase with QSHI in the upper gap occurs for $\varepsilon < -0.17$ in BI-strained Lieb [Fig.~\textcolor{blue}{S11}(a$_3$)], $\varepsilon < -0.16$ in BI-strained transition [Fig.~\textcolor{blue}{S11}(b$_3$)], $\varepsilon < -0.27$ in BI-strained Kagome [Fig.~\textcolor{blue}{S11}(c$_3$)] and for $\varepsilon < -0.21$ in UY-strained Kagome [Fig.~\textcolor{blue}{S11}(c$_2$)]. Conversely, when $\lambda_{AB} = \lambda_{BC} = 0$, this phase occurs for $\varepsilon < -0.17$ in BI-strained Lieb [Fig.~\textcolor{blue}{S11}(d$_3$)], for $\varepsilon < -0.16$ in BI-strained transition [Fig.~\textcolor{blue}{S11}(e$_3$)], for $\varepsilon < -0.23$ in BI-strained Kagome [Fig.~\textcolor{blue}{S11}(f$_3$)], and for $\varepsilon < -0.09$ in UY-strained Kagome [Fig.~\textcolor{blue}{S11}(f$_2$)]. When comparing the phase diagrams obtained with $\lambda_{AB} = \lambda_{BC} = 0$ with those considering only NNN ISO coupling, we conclude that the aforementioned phase continues to occur in the same region $\varepsilon < -0.17$ in Lieb [Fig.~\textcolor{blue}{S12}(c)], while it occurs in Kagome lattices under UY [Fig.~\textcolor{blue}{S12}(h)] and BI [Fig.~\textcolor{blue}{S12} (i)] strain in different regions, $\varepsilon^{UY}<-0.15$ and $\varepsilon^{BI}<-0.3$, respectively. Interestingly, when admitting only NN ISO coupling in Kagome lattice, the semimetallic phase with QSHI in the upper gap ceases to exist [Figs.~\textcolor{blue}{S12}(n) and \textcolor{blue}{S12}(o)], compared to the corresponding case of $\lambda_{AB} \neq 0$ and $\lambda_{BC} \neq 0$ [Figs.~\textcolor{blue}{S11}(c$_2$) and \textcolor{blue}{S11}(c$_3$)]. This shows that the NNN ISO coupling terms are responsible for this non-topological phase transition in the Kagome lattice.
 
\section{Conclusions}\label{Sec.V}

In summary, we systematically studied the TPTs driven by strain on the Lieb-Kagome lattices with ISO coupling, based on a recently proposed tight-binding Hamiltonian reported in Ref.~[\onlinecite{lima2023}] that takes into account uniaxial strains and shear applied in different crystallographic directions of the Lieb-Kagome lattice defined in terms of the interconvertibility discovered in Ref.~[\onlinecite{tony2019}] between the Lieb and Kagome lattices by defining a transition lattice that maps such structures by one control parameter. For this purpose, using the concept of a strained generic lattice,\cite{tony2019,lim_2019,osti_1527138, Cui2019,lima2023} we derived a more general Hamiltonian including the ISO coupling.

Initially, we focused our study on understanding the emergence of two band gaps in the energy spectrum of non-strained Lieb-Kagome lattices, particularly related to the introduction of ISO coupling. We demonstrated that NN ISO coupling alone does not induce band gaps in the Lieb lattice. Instead, it is the incorporation of NNN ISO coupling that disrupts the $D_{4h}$ symmetry, leading to the formation of a nearly-flat band with Chern number equal to zero.\cite{tony2019,hwang2021flat} In both transition and Kagome lattices, band gaps are produced by NN ISO coupling terms, NNN ISO coupling terms, or a combination of both. Specifically, in the Kagome lattice, the emergence of a Chern nearly-flat band with Chern number equal to $\pm1$ is observed. The process of these band gaps closing and then reopening is indicative of TPTs, confirmed by changes in the Spin Chern number and the Berry curvature, each calculated for the three energy bands.

The first TPT we identified occurred in the non-strained Lieb-Kagome lattice, specifically during the interconvertibility process at the point $\varepsilon = 118.3^\circ$. This transition point is marked by the NN ISO coupling terms $\lambda_{AB}$ and $\lambda_{BC}$ becoming null. Uniquely, this TPT represents a singular TPT driven by hypothetical strains, where the hopping and ISO coupling parameters remain strain-invariant. This finding underscores that the phase transitions induced by strain are not solely due to geometric alterations in the lattice. Instead, they are primarily driven by changes in the Hamiltonian parameters, $t_{ij}$ and $\lambda_{ij}$, as governed by Eqs.~\eqref{hopping} and \eqref{eqlambdaISO}, respectively.

Overall, TPTs were identified when varying the $\lambda$ or $n$ parameters, as well as upon applying the six types of strain to the Lieb-Kagome lattices. TPTs connecting two topological insulator phases occur when both full-band gaps, along with their corresponding local-band gaps, close and then reopen at the same $\varepsilon_c$. If the lower full-band gap closes and does not reopen, while the corresponding local-band gap remains non-zero, a negative indirect gap is observed. In this scenario, the system undergoes a non-TPT into a semimetallic regime at 1/3 filling, characterized by partially filled bands and non-quantized spin Hall conductivity. At 2/3 filling, the system behaves as an insulator, with the spin Hall conductivity carried by the helical edge states becoming experimentally observable.\cite{belgeling2012} Conversely, if both full-band gaps close and do not reopen, the system transitions to a metallic state, inhibiting any spin Hall conductivity.

The difference in the Chern numbers of the phases on either side of the TPTs identified in this work were \(\Delta C^{\uparrow} = \pm \left[(1,0,-1) - (-1,2,-1)\right] = \pm(2,-2,0)\) or \(\Delta C^{\uparrow} = \pm \left[(1,0,-1) - (-1,0,1)\right] = \pm(2,0,-2)\). We observe that the change in the Chern numbers of the two touching bands is \(\pm2\). Consequently, in all configurations with open gaps, the system behaves as a QSH insulator with spin Hall conductivity equal to \(\sigma_{SH}=\pm 1\), implying a Berry phase of \(4\pi\) associated with this touching point, in accordance with Ref.~[\onlinecite{belgeling2012}]. The Berry curvature for each band changes sign when the corresponding Chern number transitions from positive to negative, or vice versa. Remarkably, the total number of Berry curvature peaks per band is strain invariant within the first Brillouin zone, despite changes in their shape and distribution. In general, the Berry curvatures corresponding to the lower and/or upper bands exhibit two positive peaks when $C^{\uparrow}_{\text{lower}}=+1$ and/or $C^{\uparrow}_{\text{upper}}=+1$, while they show two negative peaks when $C^{\uparrow}_{\text{lower}}=-1$ or $C^{\uparrow}_{\text{upper}}=-1$. The Berry curvature associated with the middle band displays two positive and two negative peaks when $C^{\uparrow}_{\text{middle}}=0$, which transform into four positive peaks when $C^{\uparrow}_{\text{middle}}=+2$.

We believe that a systematic study, pertinent to the effects of different types of strains applied in 2D lattices, as demonstrated in the TPTs of Lieb-Kagome lattices with ISO coupling, is crucial for advancing the field of 2D topological insulators. This research highlights the potential of strain engineering to fine-tune electrical and optical properties, providing new ways for controlling band gap tunability and band deformations. Such advancements are highly relevant to the future of 2D lattice-based device technologies, making a significant contribution to both the fundamental understanding and practical applications of these flatband materials in nanoscale electronics and photonics.

\section*{Acknowledgments}

The authors would like to thank the National Council of Scientific and Technological Development (CNPq) through Universal and PQ programs and the Coordination for the Improvement of Higher Education Personnel (CAPES) of Brazil for their financial support. D.R.C gratefully acknowledges the support from CNPq grants $313211/2021-3$, $437067/2018-1$, $423423/2021-5$, $408144/2022-0$, the Research Foundation—Flanders (FWO - Vl), and the Fundação Cearense de Apoio ao Desenvolvimento Científico e Tecnológico (FUNCAP). We also thank S. H. R. Sena for helpful comments and discussions.

\appendix

\bibliography{apssamp}
\end{document}